\documentclass[12pt,english]{article}
\usepackage{times}
\usepackage[T1]{fontenc}
\usepackage{geometry}
\usepackage{rotating}
\usepackage{graphicx}
\usepackage{setspace}
\usepackage{amssymb}

\makeatletter

\providecommand{\tabularnewline}{\\}

\usepackage{bm}

\usepackage{babel}
\makeatother
\begin{document}

\section*{\noindent Time-Modulated EM Skins for Integrated Sensing and Communications}

\noindent {\footnotesize ~}{\footnotesize \par}

\noindent \vfill

\noindent L. Poli$^{(1)}$, A. Bansal$^{(2)}$, G. Oliveri$^{(1)(3)}$,
A. A. Salas-Sanchez$^{(1)(3)}$, W. Whittow$^{(2)}$, and A. Massa$^{(1)(3)(4)(5)(6)}$

\noindent \vfill

\noindent {\footnotesize ~}{\footnotesize \par}

\noindent {\scriptsize $^{(1)}$} \emph{\scriptsize CNIT - \char`\"{}University
of Trento\char`\"{} ELEDIA Research Unit }{\scriptsize \par}

\noindent {\scriptsize Via Sommarive 9, 38123 Trento - Italy}{\scriptsize \par}

\noindent {\scriptsize Website:} \emph{\scriptsize www.eledia.org/eledia-unitn}{\scriptsize \par}

\noindent {\scriptsize ~}{\scriptsize \par}

\noindent {\scriptsize $^{(2)}$ Wolfson School of Mechanical, Electrical
and Manufacturing Engineering, Loughborough University}{\scriptsize \par}

\noindent {\scriptsize Epinal Way, LE11 3TU Loughborough, United Kingdom }{\scriptsize \par}

\noindent {\scriptsize E-mail: \{A.Bansal}\emph{\scriptsize ,} {\scriptsize W.G.Whittow\}}\emph{\scriptsize @lboro.ac.uk}{\scriptsize \par}

\noindent {\scriptsize Website:} \emph{\scriptsize www.lboro.ac.uk/schools/meme/}{\scriptsize \par}

\noindent {\scriptsize ~}{\scriptsize \par}

\noindent {\scriptsize $^{(3)}$} \emph{\scriptsize ELEDIA Research
Center} {\scriptsize (}\emph{\scriptsize ELEDIA}{\scriptsize @}\emph{\scriptsize UniTN}
{\scriptsize - University of Trento)}{\scriptsize \par}

\noindent {\scriptsize DICAM - Department of Civil, Environmental,
and Mechanical Engineering}{\scriptsize \par}

\noindent {\scriptsize Via Mesiano 77, 38123 Trento - Italy}{\scriptsize \par}

\noindent \textit{\emph{\scriptsize E-mail:}} {\scriptsize \{lorenzo.poli,
giacomo.oliveri, aaron.salassanchez, andrea.massa\}@}\emph{\scriptsize unitn.it}{\scriptsize \par}

\noindent {\scriptsize Website:} \emph{\scriptsize www.eledia.org/eledia-unitn}{\scriptsize \par}

\noindent {\scriptsize ~}{\scriptsize \par}

\noindent {\scriptsize $^{(4)}$} \emph{\scriptsize ELEDIA Research
Center} {\scriptsize (}\emph{\scriptsize ELEDIA}{\scriptsize @}\emph{\scriptsize UESTC}
{\scriptsize - UESTC)}{\scriptsize \par}

\noindent {\scriptsize School of Electronic Science and Engineering,
Chengdu 611731 - China}{\scriptsize \par}

\noindent \textit{\emph{\scriptsize E-mail:}} \emph{\scriptsize andrea.massa@uestc.edu.cn}{\scriptsize \par}

\noindent {\scriptsize Website:} \emph{\scriptsize www.eledia.org/eledia}{\scriptsize -}\emph{\scriptsize uestc}{\scriptsize \par}

\noindent {\scriptsize ~}{\scriptsize \par}

\noindent {\scriptsize $^{(5)}$} \emph{\scriptsize ELEDIA Research
Center} {\scriptsize (}\emph{\scriptsize ELEDIA@TSINGHUA} {\scriptsize -
Tsinghua University)}{\scriptsize \par}

\noindent {\scriptsize 30 Shuangqing Rd, 100084 Haidian, Beijing -
China}{\scriptsize \par}

\noindent {\scriptsize E-mail:} \emph{\scriptsize andrea.massa@tsinghua.edu.cn}{\scriptsize \par}

\noindent {\scriptsize Website:} \emph{\scriptsize www.eledia.org/eledia-tsinghua}{\scriptsize \par}

\noindent {\scriptsize ~}{\scriptsize \par}

\noindent {\scriptsize $^{(6)}$} \emph{\scriptsize }{\scriptsize School
of Electrical Engineering}{\scriptsize \par}

\noindent {\scriptsize Tel Aviv University, Tel Aviv 69978 - Israel}{\scriptsize \par}

\noindent \textit{\emph{\scriptsize E-mail:}} \emph{\scriptsize andrea.massa@eng.tau.ac.il}{\scriptsize \par}

\noindent {\scriptsize Website:} \emph{\scriptsize https://engineering.tau.ac.il/}{\scriptsize \par}

\noindent \vfill

\noindent \emph{This work has been submitted to the IEEE for possible
publication. Copyright may be transferred without notice, after which
this version may no longer be accessible.}

\noindent \vfill

\newpage
\section*{Time-Modulated EM Skins for Integrated Sensing and Communications }

~

~

~

\begin{flushleft}L. Poli, A. Bansal, G. Oliveri, A. A. Salas-Sanchez,
W. Whittow, and A. Massa\end{flushleft}

\vfill

\begin{abstract}
\noindent An innovative solution, based on the exploitation of the
harmonic beams generated by time-modulated electromagnetic skins (\emph{TM-EMS}s),
is proposed for the implementation of integrated sensing and communication
(\emph{ISAC}) functionalities in a Smart Electromagnetic Environment
(\emph{SEME}) scenario. More in detail, the field radiated by a user
terminal, located at an unknown position, is assumed to illuminate
a passive \emph{TM-EMS} that, thanks to a suitable modulation of the
local reflection coefficients at the meta-atom level of the \emph{EMS}
surface, simultaneously reflects towards a receiving base station
(\emph{BS}) a {}``sum'' beam and a {}``difference'' one at slightly
different frequencies. By processing the received signals and exploiting
monopulse radar tracking concepts, the \emph{BS} both localizes the
user terminal and, as a by-product, establishes a communication link
with it by leveraging on the {}``sum'' reflected beam. Towards this
purpose, the arising harmonic beam control problem is reformulated
as a global optimization one, which is successively solved by means
of an evolutionary iterative approach to determine the desired \emph{TM-EMS}
modulation sequence. The results from selected numerical and experimental
tests are reported to assess the effectiveness and the reliability
of the proposed approach.

\vfill
\end{abstract}
\noindent \textbf{Key words}: Reconfigurable Passive \emph{EM} Skins;
Smart Electromagnetic Environment; Next-Generation Communications;
Integrated Sensing and Communications; Space-Time Coding.

\newpage
\section{Introduction and Rationale\label{sec:Introduction}}

\noindent In the last few years, the concept of Smart Electromagnetic
Environment (\emph{SEME}) has emerged as a revolutionary paradigm
to address several major challenges in next-generation wireless communications
systems \cite{Yang 2022}-\cite{Di Renzo 2019}. The \emph{SEME} vision
is based on the idea that the wave propagation properties of outdoor/indoor
environments can be tailored to enhance the {}``quality'' of the
resulting wireless links \cite{Yang 2022}-\cite{Di Renzo 2019}.
Towards this end, several technological solutions have been proposed
including both static and reconfigurable electromagnetic skins (\emph{EMS}s)
\cite{Yang 2022}\cite{Oliveri 2021c}\cite{Oliveri 2022e}. These
latter have been exploited not only for communications, but also for
the localization/tracking of users to yield improved quality-of-service
(\emph{QoS}) levels as well as to enable advanced wireless planning/deployment
\cite{Zhang 2022}\cite{Hu 2020}. As a matter of fact, the use of
\emph{EMS} technologies for sensing purposes is widely explored as
proved by the long list of recent contributions (see \cite{Zhang 2022}-\cite{Wan 2021}
and the references therein).

\noindent It is worth pointing out that the possibility to jointly
localize and establish a connection with wireless terminals not in
line-of-sight (\emph{NLOS}) is of great practical interest, especially
at higher frequencies, owing to the harsh path loss conditions \cite{Ma 2023}-\cite{Chowdhury 2020}.
This can be accomplished through dedicated \emph{HW}s/\emph{SW}s that
separately implement the two functionalities \cite{Sturm 2011}. Obviously,
a more efficient integrated sensing and communications (\emph{ISAC})
solution is preferable for costs and scalability reasons \cite{Ma 2023}-\cite{Sturm 2011}.
Towards this end, an efficient strategy could be that of exploiting
the \emph{monopulse radar} concept \cite{Manica 2008}\cite{Rocca 2009}\cite{Manica 2009}\cite{Poli 2011.b}
within the \emph{SEME} scenario\@. In principle, a traditional monopulse
radar is a multi-beam \emph{sensing} system devoted to identify the
position of a target \cite{Manica 2008}\cite{Rocca 2009}\cite{Manica 2009}.
Such a localization functionality is physically implemented by realizing
two separate simultaneous wireless beams: a difference beam, $\Delta$,
and a sum beam, $\Sigma$ \cite{Manica 2008}\cite{Rocca 2009}\cite{Manica 2009}.
Dealing with an \emph{ISAC} system, the sum beam can be used not only
for localization purposes, but also to establish a \emph{communication}
channel with the user terminal.

\noindent Adopting the monopulse radar concept in an \emph{ISAC NLOS}
scenario needs (\emph{i}) the terminal to operate as the primary source
and not like in radar scenarios where the target is passive, (\emph{ii})
the \emph{EMS} to reflect the impinging wave from the terminal towards
the \emph{BS}, and (\emph{iii}) the \emph{BS} to distinguish the received
power associated to each $\Sigma/\Delta$ beam. Unfortunately, the
practical implementation of such an idea is not trivial. As a matter
of fact, a straightforward realization of the monopulse functionality
requires the receiving antenna (i.e., the \emph{BS} in the \emph{SEME}
case) to be equipped with at least two separate adaptive feeding networks
for affording the $\Sigma/\Delta$ beams \cite{Ma 2023} as in traditional
radar monopulse systems. Following this line of reasoning, reconfigurable
\emph{EMS}s have been used in combination with multiple antennas to
build monopulse \emph{ISAC} systems \cite{Ma 2023} albeit requiring
significant modifications to the \emph{BS} hardware since independent
directive antennas and a comparator feed network must be added \cite{Ma 2023}.
Alternatively, reconfigurable \emph{EMS}s have been designed to synthesize
either a $\Sigma$ beam or a $\Delta$ beam directed towards the \emph{BS}
at different time instants \cite{Wan 2021}, but losing the simultaneous
beams generation \cite{Wan 2021}. Otherwise, simultaneous $\Sigma/\Delta$
beams could be radiated by recurring to the polarization diversity
subject to the condition that the user terminal generates dual-polarized
signals.

\noindent To overcome such limitations, this work proposes a different
approach that meets the following guidelines: to enable the simultaneous
synthesis of the $\Sigma$/$\Delta$ beams, while avoiding major hardware
modifications to both the \emph{BS} and the user terminal, by borrowing
some key concepts from Time Modulated Arrays (\emph{TMA}s) engineering
\cite{Yang 2006}\cite{Poli 2011.b}-\cite{Rocca 2019}. Let us remember
that a \emph{TMA} is an array where the antenna excitations are periodically
turned \emph{on} and \emph{off} according to a user-defined time sequence
by means of a set of radio-frequency switches \cite{Rocca 2019}.
Thanks to the \emph{time-modulation} process, a set of harmonic signals
is generated around the carrier frequency \cite{Rocca 2019}. Each
harmonic signal is associated to a unique radiation pattern, which
is controlled by the array geometry and the modulation sequence \cite{Rocca 2019}.
Such a working mechanism enables a wide variety of functionalities
in wireless sensing and communications \cite{Rocca 2019}. As for
monopulse radars, \emph{TMA}s have been previously and successfully
adopted to implement $\Sigma$/$\Delta$ beams at different harmonic
frequencies \cite{Poli 2011.b}\cite{Rocca 2019}-\cite{Rocca 2012.a}.

\noindent To translate similar concepts in the \emph{SEME} context,
let us consider time-modulated \emph{EMS}s (\emph{TM-EMS}s) as the
corresponding counterpart of \emph{TMA}s. A \emph{TM-EMS} is an artificial
passive surface where, instead of modulating the excitations of the
array elements as in \emph{TMA}s, the reflection properties of the
\emph{EMS} meta-atoms are dynamically modulated (Fig. 1) \cite{Rocca 2019}\cite{Zhang 2018}.
By leveraging the properties of time-modulated devices, advanced functionalities
in terms of reflected harmonic beam generation and control may be
synthesized in \emph{TM-EMS}s, as well. Therefore, the development
of an innovative solution, based on \emph{TM-EMS}s, for the \emph{ISAC}
scenario at hand is discussed hereinafter. Figure 1 sketches the operation
principle of the proposed \emph{ISAC TM-EMS} system. The user terminal,
which is in \emph{NLSO} with the \emph{BS}, acts as the source that
illuminates the passive \emph{TM-EMS}. Such a \emph{TM-EMS}, thanks
to the modulation in time of the surface reflection properties at
the meta-atom level, reflects towards the \emph{BS} both a $\Sigma$
beam at the carrier frequency and a $\Delta$ beam at the first harmonic
frequency. The \emph{BS} separately collects the power at the carrier
frequency, $P_{\Sigma}$, and at the first harmonic term, $P_{\Delta}$,
to compute the power index $\xi$ ($\xi\triangleq\frac{P_{\Sigma}}{P_{\Delta}}$).
According to the monopulse radar rules, the maximization of $\xi$
allows the \emph{BS} to detect the angular position of the user terminal
({}``sensing'') and, as a byproduct, to establish with him a maximum-gain
link ({}``communication'') thanks to the $\Sigma$ beam. 

\noindent Despite the similarities between \emph{TMA}s and \emph{TM-EMS}s,
there are several methodological and practical challenges to be addressed
before releasing a reliable \emph{ISAC} \emph{TM-EMS} system. For
instance, unlike \emph{TMA}s, the modulation enforced at the meta-atom
level is not {}``on-off'', but it rather involves a change in the
local reflection coefficient on the \emph{EMS} area. Moreover, while
the feed network of a \emph{TMA} controls the input signal to each
antenna, a \emph{TM-EMS} must accommodate for any impinging wave regardless
of its incidence angle.

\noindent The main innovative contributions of this work then include:
(\emph{i}) the customisation of the \emph{TMA} concepts to the control
of time-reconfigurable passive \emph{EMS}s, (\emph{ii}) the proof
of the feasibility of \emph{ISAC} solutions based on \emph{TM-EMS}s
and monopulse radar principles without major modifications on the
structure of \emph{BS}s/user-terminals, and (\emph{iii}) the assessment
of the proposed \emph{ISAC} system with full-wave numerical simulations
as well as experimental measurements on a prototype.

\noindent The outline of the paper is as follows. After the description
of the proposed \emph{ISAC} system, the design problem at hand is
formulated and an iterative procedure for the synthesis of the time-modulation
sequence of the \emph{TM-EMS} is detailed (Sect. \ref{sec:Problem-Formulation}).
A set of representative numerical and experimental results is then
reported to illustrate the features of the proposed approach as well
as to assess its effectiveness in different operative conditions (Sect.
\ref{sec:Results}). Finally, some conclusions and final remarks follow
(Sect. \ref{sec:Conclusions-and-Remarks}).

\section{\noindent Problem Formulation and Time-Modulated \emph{EMS} Design
for \emph{ISAC}\label{sec:Problem-Formulation} }

\noindent Let us consider the scenario in Fig. 1 where a rectangular
\emph{TM-EMS}, composed of $P\times Q$ time-modulated meta-atoms
and occupying an area $\Omega$, is illuminated by a time-harmonic
incident field $\mathbf{E}^{inc}\left(\mathbf{r}\right)$, generated
by the user terminal, which is modeled as a locally plane wave with
wave vector $\mathbf{k}^{inc}\left(\mathbf{r}\right)$, $\mathbf{r}=\left(x,y,z\right)$
being the position vector in a Cartesian coordinate system centered
on the \emph{EMS} aperture. The instantaneous local reflection tensor
$\overline{\overline{\Gamma}}_{pq}\left(t\right)$ at the $pq$-th
($p=1,...,P$; $q=1,...,Q$) unit cell of the \emph{TM-EMS} is assumed
to be time-modulated by means of a digital signal $U_{pq}\left(t\right)$
of period $T$ \cite{Rocca 2019}. By denoting with $t_{pq}^{on}$
($0\le t_{pq}^{on}\le T$) and $t_{pq}^{off}$ ($0\le t_{pq}^{off}\le T$)
the switch-on and switch-off instants, respectively, $U_{pq}\left(t\right)$
is mathematically described as a rectangular pulse function with values
$U_{pq}\left(t\right)=1$ when $t_{pq}^{on}\le t\le t_{pq}^{off}$
and $U_{pq}\left(t\right)=0$ otherwise. Accordingly,\begin{equation}
\overline{\overline{\Gamma}}_{pq}\left(t\right)=\overline{\overline{\Gamma}}^{on}U_{pq}\left(t\right)+\overline{\overline{\Gamma}}^{off}\widetilde{U}_{pq}\left(t\right)\label{eq:Gamma PQ nel tempo}\end{equation}
where $\widetilde{U}_{pq}\left(t\right)\triangleq1-U_{pq}\left(t\right)$
and\begin{equation}
\overline{\overline{\Gamma}}^{on/off}=\left[\begin{array}{cc}
\Gamma_{TE}^{on/off} & \Gamma_{TE-TM}^{on/off}\\
\Gamma_{TM-TE}^{on/off} & \Gamma_{TM}^{on/off}\end{array}\right]\label{eq:  unit cell response}\end{equation}
is the on/off numerically-computed reflection tensor state, whose
on-diagonal entries (i.e., $\Gamma_{TE}^{on/off}$ and $\Gamma_{TM}^{on/off}$)
are the $TE/TM$ co-polar local reflection coefficients when the cell
in the \emph{on/off} state, while the off-diagonal ones (i.e., $\Gamma_{TE-TM}^{on/off}$
and $\Gamma_{TM-TE}^{on/off}$) are the corresponding cross-polar
terms \cite{Oliveri 2022e}\cite{Yang 2019}.

\noindent Under the local periodicity approximation \cite{Oliveri 2021c},
the electric/magnetic field in a generic point of the \emph{TM-EMS}
($\mathbf{r}\in\Omega$) is given by the following expression \cite{Oliveri 2021c}\cite{Cuesta 2018}\cite{Yang 2019}\begin{equation}
\left\{ \begin{array}{l}
\mathbf{E}_{-}\left(\mathbf{r},t\right)=\sum_{p=1}^{P}\sum_{q=1}^{Q}\overline{\overline{\Gamma}}_{pq}\left(t\right)\cdot\mathbf{E}^{inc}\left(\mathbf{r}_{pq}\right)\Pi_{pq}\left(\mathbf{r}\right)\\
\mathbf{H}_{-}\left(\mathbf{r},t\right)=\frac{1}{\eta_{0}}\mathbf{k}^{inc}\left(\mathbf{r}\right)\times\mathbf{E}_{-}\left(\mathbf{r},t\right)\end{array}\right.\label{eq:E_H_meno}\end{equation}
where $\eta_{0}$ is the free-space impedance, $\mathbf{r}_{pq}$
is the barycenter of the $pq$-th ($p=1,...,P$; $q=1,...,Q$) meta-atom,
and $\Pi_{pq}\left(\mathbf{r}\right)$ is the $pq$-th pixel basis
function. 

\noindent By exploiting the Love's equivalence principle, the equivalent
electric/magnetic current induced on the \emph{EMS} ($\mathbf{r}\in\Omega$)
can be computed as follows \cite{Oliveri 2021c}\cite{Balanis 1989}\begin{equation}
\left\{ \begin{array}{l}
\mathbf{J}_{e}\left(\mathbf{r},t\right)=\widehat{\mathbf{z}}\times\mathbf{H}_{-}\left(\mathbf{r},t\right)\\
\mathbf{J}_{m}\left(\mathbf{r},t\right)=-\widehat{\mathbf{z}}\times\mathbf{E}_{-}\left(\mathbf{r},t\right)\end{array}\right.\label{eq:currents}\end{equation}
and the far-field expression of the field reflected by the \emph{TM-EMS}
($\mathbf{r}\notin\Omega$) at the instant $t$ turns out to be \cite{Oliveri 2021c}\cite{Rocca 2019}\begin{equation}
\mathbf{E}^{ref}\left(\mathbf{r},t\right)=\frac{jk_{0}}{4\pi}\frac{\exp\left(-jk_{0}r\right)\exp\left(j\omega_{0}t\right)}{r}\int_{\Omega}\widehat{\mathbf{z}}\times\left[\eta_{0}\widehat{\mathbf{z}}\times\mathbf{J}_{e}\left(\mathbf{r}',t\right)+\mathbf{J}_{m}\left(\mathbf{r}',t\right)\right]\exp\left(jk_{0}\widehat{\mathbf{r}}\cdot\mathbf{r}'\right)\mathrm{d}\mathbf{r}'\label{eq:reflected-field}\end{equation}
where $k_{0}$ is the wavenumber at the carrier frequency $f_{0}$
and $\omega_{0}$ ($\omega_{0}\triangleq2\pi f_{0}$) is the angular
frequency of the carrier.

\noindent Because of the pulse periodicity, each $pq$-th ($p=1,...,P$;
$q=1,...,Q$) instantaneous local reflection tensor can be expanded
in a Fourier series \cite{Rocca 2019}\begin{equation}
\overline{\overline{\Gamma}}_{pq}\left(t\right)=\sum_{h=-\infty}^{\infty}\overline{\overline{\Gamma}}_{pq}^{h}\exp\left(jh\frac{2\pi}{T}t\right)\label{eq:espansione gamma}\end{equation}
where $\overline{\overline{\Gamma}}_{pq}^{h}$ {[}$\overline{\overline{\Gamma}}_{pq}^{h}\triangleq\left(\overline{\overline{\Gamma}}^{on}u_{pq}^{h}+\overline{\overline{\Gamma}}^{off}\widetilde{u}_{pq}^{h}\right)${]}
is the $h$-th harmonic equivalent surface reflection tensor in the
$pq$-th ($p=1,...,P$; $q=1,...,Q$) meta-atom, while $u_{pq}^{h}\triangleq\frac{1}{T}\int_{-\frac{T}{2}}^{\frac{T}{2}}U_{pq}\left(t\right)\exp\left(-jh\frac{2\pi}{T}t\right)\mathrm{d}t$
and $\widetilde{u}_{pq}^{h}\triangleq\frac{1}{T}\int_{-\frac{T}{2}}^{\frac{T}{2}}\widetilde{U}_{pq}\left(t\right)\exp\left(-jh\frac{2\pi}{T}t\right)\mathrm{d}t$.

\noindent By substituting (\ref{eq:espansione gamma}) in (\ref{eq:E_H_meno}),
one obtains that\begin{equation}
\left\{ \begin{array}{l}
\mathbf{E}_{-}\left(\mathbf{r},t\right)=\sum_{h=-\infty}^{\infty}\mathbf{E}_{-}^{h}\exp\left(jh\frac{2\pi}{T}t\right)\\
\mathbf{H}_{-}\left(\mathbf{r},t\right)=\sum_{h=-\infty}^{\infty}\frac{1}{\eta_{0}}\mathbf{k}^{inc}\left(\mathbf{r}\right)\times\mathbf{E}_{-}^{h}\exp\left(jh\frac{2\pi}{T}t\right)\end{array}\right.\label{eq:}\end{equation}
where \begin{equation}
\mathbf{E}_{-}^{h}\left(\mathbf{r}\right)=\sum_{p=1}^{P}\sum_{q=1}^{Q}\overline{\overline{\Gamma}}_{pq}^{h}\cdot\mathbf{E}^{inc}\left(\mathbf{r}_{pq}\right)\Pi_{pq}\left(\mathbf{r}\right).\label{eq:}\end{equation}
Thus, the expression (\ref{eq:currents}) can be rewritten in the
following harmonic form\begin{equation}
\left\{ \begin{array}{l}
\mathbf{J}_{e}\left(\mathbf{r},t\right)=\sum_{h=-\infty}^{\infty}\mathbf{J}_{e}^{h}\left(\mathbf{r}\right)\exp\left(jh\frac{2\pi}{T}t\right)\\
\mathbf{J}_{m}\left(\mathbf{r},t\right)=\sum_{h=-\infty}^{\infty}\mathbf{J}_{m}^{h}\left(\mathbf{r}\right)\exp\left(jh\frac{2\pi}{T}t\right)\end{array}\right.\label{eq:correnti espanse}\end{equation}
where $\mathbf{J}_{e}^{h}\left(\mathbf{r}\right)=\widehat{\mathbf{z}}\times\frac{1}{\eta_{0}}\mathbf{k}^{inc}\left(\mathbf{r}\right)\times\mathbf{E}_{-}^{h}\left(\mathbf{r}\right)$
and $\mathbf{J}_{m}^{h}\left(\mathbf{r}\right)=-\widehat{\mathbf{z}}\times\mathbf{E}_{-}^{h}\left(\mathbf{r}\right)$.

\noindent By replacing (\ref{eq:correnti espanse}) in (\ref{eq:reflected-field}),
the harmonic expansion of the reflected field ($\mathbf{r}\notin\Omega$)
turns out to be\begin{equation}
\mathbf{E}^{ref}\left(\mathbf{r},t\right)=\sum_{h=-\infty}^{\infty}\mathbf{E}^{h}\left(\mathbf{r}\right)\exp\left[j\left(\omega_{0}+h\frac{2\pi}{T}\right)t\right],\label{eq:field totale expanded}\end{equation}
$\mathbf{E}^{h}\left(\mathbf{r}\right)$ being the $h$-th harmonic
term of the reflected field given by\begin{equation}
\begin{array}{l}
\mathbf{E}^{h}\left(\mathbf{r}\right)=\frac{jk_{0}}{4\pi}\frac{\exp\left(-jk_{0}r\right)}{r}\widehat{\mathbf{z}}\times\widehat{\mathbf{z}}\times\sum_{p=1}^{P}\sum_{q=1}^{Q}\left(\frac{1}{T}\int_{-\frac{T}{2}}^{\frac{T}{2}}\left\{ \overline{\overline{\Gamma}}^{on}U_{pq}\left(t\right)+\overline{\overline{\Gamma}}^{off}\left[1-U_{pq}\left(t\right)\right]\right\} \exp\left(-jh\frac{2\pi}{T}t\right)\mathrm{d}t\right)\cdot\\
\quad\int_{\Omega_{pq}}\left[\widehat{\mathbf{z}}\times\mathbf{k}^{inc}\left(\mathbf{r}'\right)\times\mathbf{E}^{inc}\left(\mathbf{r}_{pq}\right)-\mathbf{E}^{inc}\left(\mathbf{r}_{pq}\right)\right]\exp\left(jk_{0}\widehat{\mathbf{r}}\cdot\mathbf{r}'\right)\mathrm{d}\mathbf{r}'\end{array}\label{eq:field harmonic}\end{equation}
where $\Omega_{pq}$ is the support of the $pq$-th ($p=1,...,P$;
$q=1,...,Q$) \emph{}unit cell of the \emph{TM-EMS}.

\noindent The expressions (\ref{eq:field totale expanded})-(\ref{eq:field harmonic})
show that, as expected, the field reflected by a \emph{TM-EMS}, $\mathbf{E}^{ref}\left(\mathbf{r},t\right)$
, $\mathbf{r}\notin\Omega$, can be written as a series of harmonic
fields, each $h$-th one, $\mathbf{E}^{h}\left(\mathbf{r},t\right)$,
being spectrally located at the $h$-th angular frequency $\omega_{h}$
{[}$\omega_{h}=\left(\omega_{0}+h\frac{2\pi}{T}\right)${]}. This
implies that the frequency of each $h$-th component of the reflected
field can be easily controlled by suitably tailoring $T$, which is
of fundamental importance in the \emph{ISAC} scenario at hand since
all the harmonic beams of interest must be measured by the receiver
(i.e., the \emph{BS}) without relevant updates with respect to the
standard one. Moreover, it is worth remarking that the power associated
to each $h$-th harmonic term/beam decreases as the harmonic index
$h$ increases, hence only the first harmonic modes can be profitably
used \cite{Rocca 2019}. This is not an issue for the proposed \emph{ISAC}
system since it is based on a \emph{TM-EMS} that reflects a $\Sigma$-beam
at the carrier frequency ($h=0$) and a $\Delta$-beam at the first
harmonic component ($h=H=1$), while all the remaining are not necessary.
Furthermore, one can infer from the above derivation (\ref{eq:field harmonic}),
that the synthesis of a desired functionality/pattern at the central
($h=0$) and harmonic ($h\neq0$) frequencies requires the definition
of suitable sequences of switch-on/switch-off instants, which are
coded into the time-modulation vector $\mathcal{T}$ ($\mathcal{T}\triangleq\left\{ t_{pq}^{on},t_{pq}^{off};\, p=1,...,P;\, q=1,...,Q\right\} $.

\noindent Therefore, after defining the cost function $\Phi\left(\mathcal{T}\right)$
to enforce the desired pattern features at the harmonic beams of interest
(i.e., $h=0,...,H$; $H=1$) as follows \cite{Poli 2011.b}\begin{equation}
\Phi\left(\mathcal{T}\right)=\sum_{h=0}^{H}\int\mathcal{R}\left[\left|\mathbf{E}^{h}\left(\mathbf{r}\right)\right|^{2}-\mathcal{U}^{h}\left(\mathbf{r}\right)\right]d\mathbf{r}+\int\mathcal{R}\left[\mathcal{L}^{h}\left(\mathbf{r}\right)-\left|\mathbf{E}^{h}\left(\mathbf{r}\right)\right|^{2}\right]d\mathbf{r}\label{eq:cost function}\end{equation}
where $\mathcal{L}^{h}\left(\mathbf{r}\right)$ and $\mathcal{U}^{h}\left(\mathbf{r}\right)$
are the lower and the upper masks for the $h$-th ($h=0,...,H$) harmonic
pattern, while $\mathcal{R}\left[\cdot\right]$ ($\mathcal{R}\left[\cdot\right]\triangleq\left[\cdot\right]\times\mathcal{H}\left[\cdot\right]$,
$\mathcal{H}\left[\cdot\right]$ being the Heaviside function) is
the ramp function, the synthesis problem at hand can be stated as
that of determining the optimal time-modulation setup, $\mathcal{T}^{*}$,
such that $\Phi\left(\mathcal{T}\right)$ is minimized\begin{equation}
\mathcal{T}*=\arg\left\{ \min_{\mathcal{T}}\left[\Phi\left(\mathcal{T}\right)\right]\right\} .\label{eq:optimization}\end{equation}
While the design of the desired $\Sigma$/$\Delta$-beams may be carried
out by defining suitable masks $\mathcal{L}^{h}\left(\mathbf{r}\right)$
and $\mathcal{U}^{h}\left(\mathbf{r}\right)$ and minimizing (\ref{eq:cost function})
with respect to $\mathcal{T}$, the efficiency of the synthesis process
can be significantly improved by adding a set of constraints on the
switch-on and switch-off time instants to take into account the nature
of the desired beams. From the \emph{TMA} theory \cite{Poli 2011.b}\cite{Rocca 2019},
it is well known that a $\Delta$-shaped first ($h=1$) harmonic beam
can be generated by setting $t_{P-p+1,q}^{on}=t_{pq}^{on}+\frac{T}{2}$
($p=1,...,\frac{P}{2}$; $q=1,...,Q$) in the digital sequences of
the $P\times Q$ \emph{TM-EMS} atoms. In other words, a half-period
shift between the left and the right portions of the \emph{EMS} aperture
automatically enforces the desired deep null in the $\Delta$ beam
\cite{Poli 2011.b}. Thanks to this, the size of $\mathcal{T}$ (i.e.,
the number of solution descriptors) is reduced from the original $P\times Q\times2$
real-valued entries to $P\times Q$. Nevertheless, the minimization
of (\ref{eq:cost function}) is still a challenging task owing to
the highly non-linear and multi-minima nature of $\Phi\left(\mathcal{T}\right)$.

\noindent To properly address the complexity of such a global optimization
problem, an evolutionary strategy, inspired by the \emph{TMA} design
\cite{Poli 2011.b} and based on the particle swarm (\emph{PS}) mechanisms
\cite{Rocca 2009w}, is adopted. More in detail, an iterative loop
is carried out ($\ell$ being the iteration index, $\ell=1,...,L$)
where at each $\ell$-th ($\ell=1,...,L$) iteration, a set of $C$
guess solutions, \{$\mathcal{T}_{c}^{\left(\ell\right)}$; $c=1,..,C$\},
is updated according to the \emph{PS} evolution mechanisms until either
a maximum number of iterations is reached (i.e., $\ell=L$) or the
stagnation condition on the optimal value of the cost function, $\mathcal{T}_{\ell}^{opt}=\arg\left\{ \min_{c=1,..,C;l=1,...,\ell}\left[\Phi\left(\mathcal{T}_{c}^{\left(l\right)}\right)\right]\right\} $,
holds true \cite{Rocca 2009w}. The optimal \emph{TM-EMS} control
sequence $\mathcal{T}*$, given by $\mathcal{T}*=\arg\left\{ \min_{c=1,..,C;\ell=1,...,L}\left[\Phi\left(\mathcal{T}_{c}^{\left(\ell\right)}\right)\right]\right\} $,
is finally outputted.

\noindent For the sake of clarity, it is worthwhile to point out the
key differences with respect to the control of a standard \emph{TMA}:
(\emph{i}) unlike a \emph{TMA}, a \emph{TM-EMS} is a passive reconfigurable
structure reflecting an external field, $\mathbf{E}^{inc}\left(\mathbf{r}\right)$,
which is not under the control of the \emph{EMS} itself, but it depends
on both the position and the type of the user terminal; (\emph{ii})
the modulation at the meta-atom level in \emph{TM-EMS}s varies the
surface reflection coefficient between different states (e.g., $\overline{\overline{\Gamma}}^{on}$/$\overline{\overline{\Gamma}}^{off}$)
instead of connecting/disconnecting each antenna from the feed network
as in \emph{TMA}s; (\emph{iii}) the aim of a \emph{TM-EMS} is to yield
the anomalous reflection functionality as well as the harmonic beam
shaping without exploiting {}``static excitations'' as usually done
in \emph{TMA}s \cite{Poli 2011.b}.

\section{\noindent Numerical and Experimental Validation\label{sec:Results}}

\noindent This section is aimed at illustrating the potentialities
of the proposed \emph{ISAC} solution as well as to assess the effectiveness
of the synthesized \emph{TM-EMS}s with numerical full-wave Ansys HFSS
\cite{HFSS 2021} simulations as well as experimental measurements
on a prototype. 

\noindent Let us model the incident field, generated by the user terminal,
as a $\varphi$-polarized plane wave at the carrier frequency $f_{0}=5.5$
GHz with $1$ {[}V/m{]} magnitude that impinges on the \emph{EMS}
aperture from the direction $\theta_{inc}$ ($\varphi_{inc}=0$).
The \emph{TM-EMS} is composed of square unit cells with $0.45\times0.45\lambda^{2}$
area and the time-modulation period has been chosen equal to $T=10^{-6}$
{[}s{]}, while the optimization of the time-modulation vector $\mathcal{T}$
has been carried out according to the state-of-the-art \emph{PS} guidelines
\cite{Poli 2011.b} by setting a swarm size of $C=20$ guess solutions,
$L=1000$ iterations, and an inertial weight equal to $0.4$, while
the social and the cognitive acceleration coefficients have been fixed
to $2.0$.

\subsection{\noindent Numerical Results}

\noindent In the first numerical experiment, the user has been assumed
to be along the direction $\theta_{inc}=40$ {[}deg{]} {[}i.e., $u_{inc}=0.64$
($u_{inc}\equiv u_{UT}$) and $v_{inc}=0.0$ ($v_{inc}\equiv v_{UT}$),
being $v\triangleq\sin\theta\sin\varphi$ and $v\triangleq\sin\theta\cos\varphi${]}
and connected to a \emph{BS} situated at an angle $\theta_{refl}=-20$
{[}deg{]} {[}i.e., $u_{refl}=0.34$ ($u_{refl}\equiv u_{BS}$) and
$v_{refl}=0.0$ ($v_{refl}\equiv v_{BS}$){]} with respect to a $P\times Q=10\times10$
\emph{TM-EMS}. The upper and lower masks for the sum and the difference
beams in (\ref{eq:cost function}) have been set as in Figs. 2(\emph{a})-2(\emph{d})
with a $-10$ {[}dB{]} sidelobe level. Moreover, the unit cell response
(\ref{eq:  unit cell response}) has been modeled as an ideally-reflecting
structure with $100\%$ efficiency (i.e., $\overline{\overline{\Gamma}}^{on}=\mathcal{I}$,
$\overline{\overline{\Gamma}}^{off}=-\mathcal{I}$, $\mathcal{I}$
being the identity matrix).

\noindent By optimizing the time-modulation vector $\mathcal{T}*$
according to the synthesis procedure detailed in Sect. \ref{sec:Problem-Formulation},
the time-modulated \emph{TM-EMS} control sequence turned out to be
characterized by the normalized pulse duration $\tau_{pq}$ ($\tau_{pq}\triangleq\frac{t_{pq}^{off}-t_{pq}^{on}}{T}$)
in Fig. 2(\emph{g}) and the normalized rising instant $\widetilde{t}_{pq}^{on}$
($\widetilde{t}_{pq}^{on}\triangleq\frac{t_{pq}^{on}}{T}$) in Fig.
2(\emph{h}). The corresponding distributions of the magnitude of the
harmonic beams radiated at the carrier, $\mathbf{E}^{0}\left(u,v\right)$,
and at the first harmonic, $\mathbf{E}^{1}\left(u,v\right)$, are
shown in Fig. 2(\emph{e}) and Fig. 2(\emph{f}), respectively. As it
can be observed, the \emph{TM-EMS} reflects the field radiated by
the user-terminal towards the \emph{BS} with a $\Sigma$-pattern at
the carrier frequency ($h=0$) {[}Fig. 2(\emph{e}){]} and a $\Delta$-pattern
at the $h=1$ harmonic {[}Fig. 2(\emph{f}){]}. In both cases, the
sidelobe distribution appears well controlled with no major secondary
lobes.

\noindent The effectiveness of the \emph{TM-EMS} design process when
varying the position of the \emph{BS} has been assessed next. The
plots of the reflected beams when $\theta_{refl}$ is varied from
the value of the first test case (i.e., $\theta_{refl}=-20$ {[}deg{]})
to $\theta_{refl}=-10$ {[}deg{]} {[}i.e., $u_{BS}=0.17$ and $v_{BS}=0.0${]}
{[}Figs. 3(\emph{a})-3(\emph{b}){]} or $\theta_{refl}=0$ {[}deg{]}
{[}i.e., $u_{BS}=v_{BS}=0.0${]} {[}Figs. 3(\emph{c})-3(\emph{d}){]}
show that the conceived \emph{ISAC} system is able to reconfigure
the reflection status of the \emph{EMS} meta-atoms to adequately accommodate
the desired $\Sigma/\Delta$-beam directions regardless of the \emph{BS}
position.

\noindent To quantify the robustness of the proposed solution, the
plot of the $\Sigma/\Delta$ beam ratio $\xi$ \cite{Poli 2011.b},\begin{equation}
\xi=\left.\frac{\left|\mathbf{E}^{0}\left(\mathbf{r}\right)\right|^{2}}{\left|\mathbf{E}^{1}\left(\mathbf{r}\right)\right|^{2}}\right\rfloor _{\theta=\theta_{refl}},\label{eq:monopulse ratio}\end{equation}
 versus $\theta_{refl}$ is reported in Fig. 4. Despite the small
\emph{EMS} aperture $\Omega$ and the unavoidable scan losses caused
its the planar nature as well as the fact that the anomalous reflection
is directed away from broadside, the value of $\xi$ ranges in the
interval $15.4\le\xi\le24.9$. Such a positive result (i.e., $\xi>10$)
points out that the \emph{TM-EMS} is not constrained to be oriented
towards the \emph{BS} for enabling a correct \emph{ISAC} working of
the system.

\noindent The third numerical experiment has been devoted to investigate
the dependence of the \emph{ISAC} performance on the location of the
user terminal. Towards this end, the position of the \emph{BS} has
been kept fixed at $\theta_{refl}=0$ {[}deg{]} {[}i.e., $u_{BS}=v_{BS}=0.0${]}
while the user direction has been varied within the range $20$ {[}deg{]}
$\le\theta_{inc}\le30$ {[}deg{]} (Fig. 5). Also in this case, the
plot of $\xi$ versus $\theta_{inc}$ {[}Fig. 5(\emph{a}){]} and the
corresponding reflected patterns {[}Figs. 5(\emph{b})-5(\emph{e}){]}
assess an effective control of the harmonic beams according to the
\emph{ISAC} requirements regardless of the angular position of the
user {[}e.g., $\Delta$-beam ($h=1$) - Fig. 5(\emph{c}) and Fig.
5(\emph{e}){]}. However the adjustment of the on/off sequences to
compensate for the different incident fields, since the user position
changes while $\theta_{refl}$ is kept unaltered, yields to a unavoidable
non-symmetric distributions of the lobes in the harmonic patterns
{[}e.g., $\Sigma$-beam ($h=0$) - Fig. 5(\emph{b}) and Fig. 5(\emph{d}){]}
even though the pattern masks in (\ref{eq:cost function}) are symmetric.
Moreover, Figure 5(\emph{a}) highlights that the resolution accuracy
reduces as $\theta_{inc}$ and $\theta_{refl}$ gets angularly closer
(i.e., $\xi=2$ when $\theta_{inc}=20$ {[}deg{]}, while $\xi=25$
when $\theta_{inc}=30$ {[}deg{]} being $\theta_{refl}=0$ {[}deg{]}). 

\noindent Previous experiments refer to an architecture that admits
an independent setup of $t_{pq}^{on}$ and $t_{pq}^{off}$ at each
$pq$-th ($p=1,...,P$; $q=1,...,Q$) \emph{EMS} unit cell, which
may yield to complex control electronics for separating biasing and
control lines of each \emph{EMS} switch. The system can be significantly
simplified if the scan direction of the reflected beam has to vary
only in the horizontal plane. Under such an hypothesis, only a column-wise
control of the \emph{TM-EMS} is needed so that $t_{pq}^{on}=t_{p}^{on}$
and $t_{pq}^{off}=t_{p}^{off}$ ($q=1,...,Q$; $p=1,...,P$).

\noindent Figure 6 is related to the same test case and \emph{EMS}
layout ($P\times Q=10\times10$) of Fig. 5, but assuming a column-wise
working of the \emph{TM-EMS}. The plot of $\xi$ in Fig. 6(\emph{a})
proves the effectiveness of the simplified \emph{ISAC} system being
$6.9\le\xi\le17.1$ when $20$ {[}deg{]} $\leq\theta_{inc}\leq40$
{[}deg{]}. For illustrative purposes, the harmonic reflected patterns,
generated by the time-modulated sequences in Fig. 7, when $\theta_{inc}=40$
{[}deg{]} (i.e., $u_{UT}=0.64$ and $v_{UT}=0.0$) and $\theta_{inc}=20$
{[}deg{]} (i.e., $u_{UT}=0.34$ and $v_{UT}=0.0$) are reported in
Figs. 6(\emph{b})-6(\emph{c}) and Figs. 6(\emph{d})-6(\emph{e}), respectively.

\noindent The process enabling a passive listening \emph{BS} to detect
the position of an unknown user by means of the conceived \emph{TM-EMS}-based
system is then illustrated. More specifically, a $P\times Q=10\times10$
column-wise time-controlled \emph{EMS} has been assumed to lie in
a scenario where there is a \emph{BS} at $\theta_{refl}=0$ that senses
the environment to localize a user terminal placed at $\theta_{inc}=40$
{[}deg{]}. Towards this end, the \emph{TM-EMS} has been dynamically
reconfigured by sequentially varying the candidate direction $\widehat{\theta}_{inc}$
within the range $0$ {[}deg{]} $\le\widehat{\theta}_{inc}\le50$
{[}deg{]}, while the (fixed) \emph{BS} has been delegated to measure
$P_{\Sigma}$ and $P_{\Delta}$ for computing the index $\xi$ in
correspondence with each $\widehat{\theta}_{inc}$ value. The plot
of $\xi$ in Fig. 8(\emph{a}) shows that, as expected, the power ratio
value exhibits a peak when $\widehat{\theta}_{inc}=\theta_{inc}$.
For completeness, the plots of the corresponding $\Sigma$/$\Delta$-beams
at representative angular samples $\widehat{\theta}_{inc}$ are reported
in Fig. 8 along with the corresponding \emph{TM-EMS} time-control
sequences. It is interesting to notice that the alignment of the $h=0$
and $h=1$ beams towards the \emph{BS} (i.e., $\theta_{rifl}=0$ $\to$
$u_{BS}=v_{BS}=0.0$) is obtained only when $\widehat{\theta}_{inc}=\theta_{inc}$
{[}Figs. 8(\emph{j})-8(\emph{k}){]} by setting $\mathcal{T}*$ as
in Figs. 8(\emph{l})-8(\emph{m}). Otherwise {[}i.e., $\widehat{\theta}_{inc}=0$
{[}deg{]} - Figs. 8(\emph{b})-8(\emph{c}); $\widehat{\theta}_{inc}=20$
{[}deg{]} - Figs. 8(\emph{f})-8(\emph{g}){]}, the maximum/null of
the sum/difference pattern is not at $u_{BS}=v_{BS}=0.0$ when $\widehat{\theta}_{inc}\neq\theta_{inc}$.

\noindent In the subsequent test case, the scalability of the proposed
\emph{ISAC} system has been verified by considering a wider \emph{EMS}
aperture of $P\times Q=24\times24$ unit cells, while setting $\theta_{inc}=30$
{[}deg{]} (i.e., $u_{UT}=0.5$ and $v_{UT}=0.0$) and $\theta_{refl}=0$
{[}deg{]} (i.e., $u_{BS}=v_{BS}=0.0$). The reflected beams synthesized,
after the \emph{PS} optimization, with the time-modulated sequence
in Figs. 9(\emph{c})-9(\emph{d}) are shown in Fig. 9(\emph{a}) ($h=0$)
and Fig. 9(\emph{b}) ($h=1$). It can be inferred that the increased
area of the \emph{EMS} $\Omega$ has been profitably exploited to
provide narrower $\Sigma/\Delta$-beams and well controlled sidelobes
{[}e.g., Fig. 9(\emph{b}) vs. Fig. 5(\emph{c}){]} notwithstanding
the simplified column-wise controlled \emph{TM-EMS}.

\noindent The final numerical experiment is aimed at validating the
proposed \emph{ISAC} scheme in the presence of a non-ideal \emph{TM-EMS}
by emulating the actual micro-scale response of the reconfigurable
passive \emph{TM-EMS} with the Ansys HFSS simulator. By considering
an FR4 ($\varepsilon_{r}=4.29$, $\tan\delta=2.0\times10^{-2}$) substrate
with $1.6\times10^{-3}$ {[}m{]} thickness and the MACOM MADP-000907-14020
diodes as switching devices, the single-bit structure in \cite{Oliveri 2022e}
has been chosen for the \emph{TM-EMS} meta-atom by also including
the biasing circuits and RF chokes (Fig. 10). The values of the geometrical
descriptors of the unit cell in Fig. 10 are provided in Tab. I. Numerically,
the \emph{TM-EMS} has been modeled by using the finite-element boundary-integral
(\emph{FE-BI}) method, without periodic boundary conditions approximations
or the use of Floquet ports, to take into account both the internal
coupling and the edge diffraction effects of the \emph{TM-EMS} layout.

\noindent The $\Sigma/\Delta$ patterns radiated by the $P\times Q=8\times8$
non-ideal \emph{TM-EMS} arrangement in Fig. 11(\emph{a}) when $\theta_{inc}=30$
{[}deg{]} (i.e., $u_{UT}=0.5$ and $v_{UT}=0.0$) and $\theta_{refl}=0$
{[}deg{]} (i.e., $u_{BS}=v_{BS}=0.0$) are shown in Figs. 11(\emph{b})-11(\emph{c}).
It turns out that the \emph{TM-EMS} reflects a well-defined sum pattern
at the $h=0$ harmonic {[}Fig. 11(\emph{b}){]} and the $h=1$ difference
beam exhibits the desired null along the $\theta_{refl}$ ($\to u_{BS}$)
direction {[}Fig. 11(\emph{c}){]}. As for the main quantization lobe
in the carrier frequency pattern {[}Fig. 11(\emph{b}){]}, it is caused
by the specular reflection of the incident beam and it can be reduced/mitigated
by adopting a more advanced \emph{EMS} meta-atom design \cite{Oliveri 2021c}\cite{Oliveri 2022e}.

\noindent Similar conclusions on the $\Sigma/\Delta$ beam control
and sidelobe distribution hold true for the full-wave simulation of
a \emph{TM-EMS} featuring a wider \emph{EMS} aperture with $P\times Q=16\times16$
unit cells {[}Fig. 12(\emph{a}){]}.

\subsection{\noindent Experimental Measurements}

\noindent \emph{A TM-EMS} prototype has then been manufactured and
measured (Fig. 13). More specifically, the optimized $P\times Q=16\times16$
design in Fig. 12(a) with a total area of $\Omega\approx40\times40$
{[}cm$^{2}${]} has been fabricated by printing the meta-atoms arrangement,
including the surface-mounted diodes, on a $1.6\times10^{-3}$ {[}m{]}-thick
FR4 substrate {[}Fig. 13(\emph{a}){]}, while realizing the biasing
lines on a separate $1.0\times10^{-4}$ {[}m{]}-thick FR4 substrate
(Tab. I) then connected to a Raspberry PI control unit {[}Fig. 13(\emph{b}){]}. 

\noindent The measurement setup is sketched in Fig. 14(\emph{a}),
while the measurement process has been carried out by using a RF source
operating at $f_{0}=5.5$ {[}GHz{]} and setting $T=3.0\times10^{-6}$
{[}s{]}. In particular, the \emph{TM-EMS} prototype has been mounted
on a wooden frame {[}Fig. 13(\emph{b}){]}, coated with a radar absorbing
material {[}Fig. 13(\emph{a}){]}, and an Ainfo LB4550 standard gain
horn antenna has been used for simulating both the source and the
field probe {[}Fig. 14(\emph{b}){]}. Moreover, the receiving antenna
has been mounted on a mechanically rotating platform to automatically
collect the reflected pattern samples in the whole azimuth plane {[}Fig.
14(\emph{b}){]}. Otherwise, the transmitting antenna has been installed
on a fixed support located $80$ {[}cm{]} away from the \emph{TM-EMS}
and with a $30$ {[}deg{]} elevation offset to avoid blockage effects
during the scan {[}Fig. 14(\emph{a}){]}.

\noindent The plots of the Ansys HFSS-simulated and measured beam
patterns at the carrier ($h=0$) frequency {[}Fig. 15(\emph{a}){]}
and at the first ($h=1$) harmonic {[}Fig. 15(\emph{b}){]} show a
satisfactory agreement between numerically-simulated and practically-measurable
\emph{TM-EMS} reflection properties at both frequencies. Such a result
also experimentally proves the feasibility of an \emph{ISAC} system
based on \emph{TM-EMS}s by also assessing its reliability even when
the target to be detected is located close to the \emph{EMS} (Fig.
14) and despite the relatively inexpensive fabrication of the prototype
featuring a high-loss substrate.

\section{\noindent Concluding Remarks\label{sec:Conclusions-and-Remarks}}

\noindent The implementation of an \emph{ISAC} system, based on the
well-established monopulse radar concepts and exploiting the dynamic
control of the harmonic beam-patterns reflected by a \emph{TM-EMS},
has been proposed. The control of the reflected beams features, obtained
from the design of the time-modulation sequences of the \emph{TM-EMS},
has been formulated as an optimization problem then solved with a
\emph{PS}-driven iterative approach.

\noindent From the numerical/experimental results, the following outcomes
and accomplishments can be drawn:

\begin{itemize}
\item \noindent the proposed approach is able to fulfill the requirements
of an \emph{ISAC} system in terms of generation of a $\Sigma$-beam
at the carrier ($h=0$) and a $\Delta$-beam at the first ($h=1$)
harmonic frequency regardless of the relative angular position of
both the user terminal (\emph{source}) and the \emph{BS} base station
(\emph{receiver}) (Figs. 4-5);
\item the use of simplified \emph{TM-EMS} architectures with a column-wise
control (Fig. 7) still guarantees an excellent $\Sigma/\Delta$-beam
control for azimuth-selective sensing and communications functionalities
(Fig. 6);
\item the feasibility of an \emph{ISAC} system based on \emph{TM-EMS}s is
confirmed by both full-wave numerical simulations (Fig. 10) and experimental
measurements carried out on a prototype built with a relatively inexpensive
HW (Fig. 14).
\end{itemize}
\noindent Future works, beyond the scope of the current manuscript,
will be aimed at generalizing the proposed \emph{ISAC} system to multi-bit
architectures as well as to include additional functionalities thanks
to a further and deeper exploitation of the non-negligible harmonic
terms of the \emph{TM-EMS} reflected field.

\section*{\noindent Acknowledgements}

\noindent This work benefited from the networking activities carried
out within the Project {}``ICSC National Centre for HPC, Big Data
and Quantum Computing (CN HPC)'' funded by the European Union - NextGenerationEU
within the PNRR Program (CUP: E63C22000970007), the Project {}``DICAM-EXC''
funded by the Italian Ministry of Education, Universities and Research
(MUR) (Departments of Excellence 2023-2027, grant L232/2016), the
Project {}``INSIDE-NEXT - Indoor Smart Illuminator for Device Energization
and Next-Generation Communications'' funded by the Italian Ministry
for Universities and Research within the PRIN 2022 Program (CUP: E53D23000990001),
and the Project {}``AURORA - Smart Materials for Ubiquitous Energy
Harvesting, Storage, and Delivery in Next Generation Sustainable Environments''
funded by the Italian Ministry for Universities and Research within
the PRIN-PNRR 2022 Program, and the Project Partnership on {}``Telecommunications
of the Future'' (PE00000001 - program {}``RESTART''), funded by
the European Union under the Italian National Recovery and Resilience
Plan (NRRP) of NextGenerationEU (CUP: E63C22002040007). A. Massa wishes
to thank E. Vico and L. Massa for the never-ending inspiration, support,
guidance, and help.

\newpage
\section*{FIGURE CAPTIONS}

\begin{itemize}
\item \textbf{Figure 1.} \emph{Problem Scenario} - Sketch of the proposed
\emph{ISAC} system.
\item \textbf{Figure 2.} \emph{Illustrative Example} (\emph{Ideal Meta-Atom},
$P\times Q=10\times10$, $\theta_{inc}=40$ {[}deg{]}, $\theta_{refl}=-20$
{[}deg{]}) - Plots of (\emph{a})(\emph{b}) the lower masks, (\emph{c})(\emph{d})
upper masks, and (\emph{e})(\emph{f}) reflected power patterns at
(\emph{a})(\emph{c})(\emph{e}) the central frequency ($h=0$) and
(\emph{b})(\emph{d})(\emph{f}) the first ($h=1$) harmonic term along
with the corresponding (\emph{g}) $\tau_{pq}$ and (\emph{h}) $\widetilde{t}_{pq}^{on}$
sets.
\item \textbf{Figure 3.} \emph{Numerical Analysis} (\emph{Ideal Meta-Atom},
$P\times Q=10\times10$, $\theta_{inc}=40$ {[}deg{]}) - Plots of
reflected power patterns at (\emph{a})(\emph{c}) the carrier ($h=0$)
and (\emph{b})(\emph{d}) the first ($h=1$) harmonic frequency when
(\emph{a})(\emph{b}) $\theta_{refl}=-10$ {[}deg{]} and (\emph{c})(\emph{d})
$\theta_{refl}=0$ {[}deg{]}.
\item \textbf{Figure 4.} \emph{Numerical Analysis} (\emph{Ideal Meta-Atom},
$P\times Q=10\times10$, $\theta_{inc}=40$ {[}deg{]}) - Plot of $\xi$
versus $\theta_{refl}$.
\item \textbf{Figure 5.} \emph{Numerical Analysis} (\emph{Ideal Meta-Atom},
$P\times Q=10\times10$, $\theta_{refl}=0$ {[}deg{]}) - Plots of
(\emph{a}) $\xi$ versus $\theta_{inc}$and of the reflected power
patterns at (\emph{b})(\emph{d}) the carrier frequency ($h=0$) and
(\emph{c})(\emph{e}) the first ($h=1$) harmonic when (\emph{b})(\emph{c})
$\theta_{inc}=30$ {[}deg{]} and (\emph{d})(\emph{e}) $\theta_{inc}=20$
{[}deg{]}.
\item \textbf{Figure 6.} \emph{Numerical Analysis} (\emph{Ideal Meta-Atom},
$P\times Q=10\times10$, $\theta_{refl}=0$ {[}deg{]}, \emph{column-wise
control}) - Plots of (\emph{a}) $\xi$ versus $\theta_{inc}$ and
of the reflected power patterns at (\emph{b})(\emph{d}) the carrier
frequency ($h=0$) and (\emph{c})(\emph{e}) the first ($h=1$) harmonic
when (\emph{b})(\emph{c}) $\theta_{inc}=40$ {[}deg{]} and (\emph{d})(\emph{e})
$\theta_{inc}=20$ {[}deg{]}.
\item \textbf{Figure 7.} \emph{Numerical Analysis} (\emph{Ideal Meta-Atom},
$P\times Q=10\times10$, $\theta_{refl}=0$ {[}deg{]}, \emph{column-wise
control}) - Plots of (\emph{a})(\emph{c}) $\tau_{pq}$ and (\emph{b})(\emph{d})
$\widetilde{t}_{pq}^{on}$ when (\emph{a})(\emph{b}) $\theta_{inc}=40$
{[}deg{]} and (\emph{c})(\emph{d}) $\theta_{inc}=20$ {[}deg{]}.
\item \textbf{Figure 8.} \emph{Numerical Analysis} (\emph{Ideal Meta-Atom},
$P\times Q=10\times10$, $\theta_{refl}=0$ {[}deg{]}, \emph{column-wise
control}, $\theta_{inc}=40$ {[}deg{]}) - Plots of (\emph{a}) $\xi$
versus $\widehat{\theta}_{inc}$ and of reflected power patterns at
(\emph{b})(\emph{f})(\emph{j}) the carrier frequency ($h=0$), (\emph{c})(\emph{g})(\emph{k})
the first ($h=1$) harmonic along with the corresponding (\emph{d})(\emph{h})(\emph{l})
$\tau_{pq}$ and (\emph{e})(\emph{i})(\emph{m}) $\widetilde{t}_{pq}^{on}$
values when (\emph{b})(\emph{c})(\emph{d})(\emph{e}) $\widehat{\theta}_{inc}=0$
{[}deg{]}, (\emph{f})(\emph{g})(\emph{h})(\emph{i}) $\widehat{\theta}_{inc}=20$
{[}deg{]}, and (\emph{j})(\emph{k})(\emph{l})(\emph{m}) $\widehat{\theta}_{inc}=40$
{[}deg{]}.
\item \textbf{Figure 9.} \emph{Numerical Analysis} (\emph{Ideal Meta-Atom},
$P\times Q=24\times24$, $\theta_{inc}=30$ {[}deg{]}, $\theta_{refl}=0$
{[}deg{]}, \emph{column-wise control}) - Plots of reflected power
patterns at (\emph{a}) the carrier ($h=0$) and (\emph{b}) the first
($h=1$) harmonic along with the corresponding (\emph{c}) $\tau_{pq}$
and (\emph{d}) $\widetilde{t}_{pq}^{on}$ values.
\item \textbf{Figure 10.} \emph{Numerical Analysis} (\emph{HFSS Full-wave
Modeling}) - Meta-atom geometry: (\emph{a}) top view, (\emph{b}) side
view, and (\emph{c}) bottom view.
\item \textbf{Figure 11.} \emph{Numerical Analysis} (\emph{HFSS Full-wave
Modeling}, $P\times Q=8\times8$, $\theta_{inc}=30$ {[}deg{]}, $\theta_{refl}=0$
{[}deg{]}, \emph{column-wise control}) - Sketch of (\emph{a}) the
full-wave model of the \emph{TM-EMS} and plots of (\emph{b})(\emph{c})
the reflected power patterns at (\emph{b}) the carrier ($h=0$) and
(\emph{c}) the first ($h=1$) harmonic.
\item \textbf{Figure 12.} \emph{Numerical Analysis} (\emph{HFSS Full-wave
Modeling}, $P\times Q=16\times16$, $\theta_{inc}=30$ {[}deg{]},
$\theta_{refl}=0$ {[}deg{]}, \emph{column-wise control}) - Sketch
of (\emph{a}) the full-wave model of the \emph{TM-EMS} architecture
and plots of (\emph{b})(\emph{c}) reflected power patterns at (\emph{b})
the carrier ($h=0$) and (\emph{c}) the first ($h=1$) harmonic.
\item \textbf{Figure 13.} \emph{Experimental Results} ($P\times Q=16\times16$,
$\theta_{inc}=30$ {[}deg{]}, $\theta_{refl}=0$ {[}deg{]}, \emph{column-wise
control}) - Photo of the (\emph{a}) front and (\emph{b}) back of the
\emph{TM-EMS} prototype as installed within the measurement setup.
\item \textbf{Figure 14.} \emph{Experimental Results} ($P\times Q=16\times16$,
$\theta_{inc}=30$ {[}deg{]}, $\theta_{refl}=0$ {[}deg{]}, \emph{column-wise
control}) - Measurement setup: (\emph{a}) logical sketch and (\emph{b})
photo.
\item \textbf{Figure 15.} \emph{Experimental Results} ($P\times Q=16\times16$,
$\theta_{inc}=30$ {[}deg{]}, $\theta_{refl}=0$ {[}deg{]}, \emph{column-wise
control}) - Plots of the measured reflected power patterns at (\emph{a})
the carrier ($h=0$) and (\emph{b}) the first ($h=1$) harmonic.
\end{itemize}

\section*{TABLE CAPTIONS}

\begin{itemize}
\item \textbf{Table I.} \emph{Numerical Validation} - Meta-atom descriptors.
\end{itemize}
\noindent ~

\newpage
\begin{center}~\vfill\end{center}

\begin{center}\includegraphics[%
  width=0.95\columnwidth]{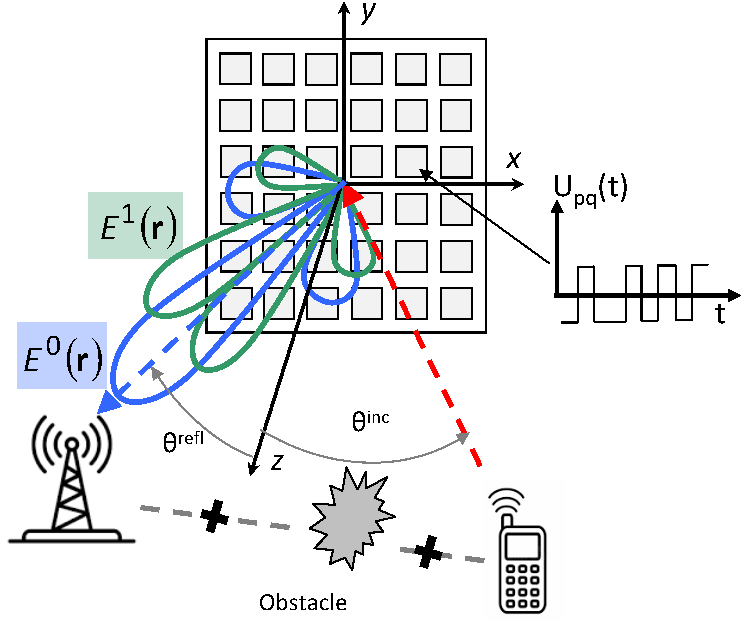}\end{center}

\begin{center}~\vfill\end{center}

\begin{center}\textbf{Fig. 1 - L. Poli} \textbf{\emph{et al.,}} {}``Time-Modulated
EM Skins for Integrated ...''\end{center}

\noindent ~

\newpage
\begin{center}\begin{tabular}{cc}
\includegraphics[%
  width=0.35\columnwidth]{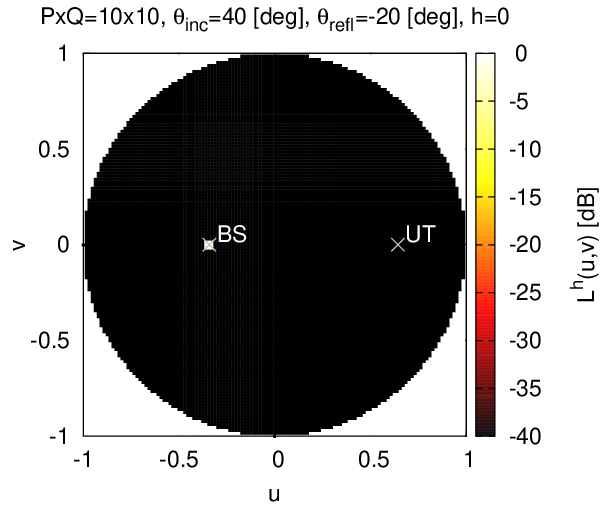}&
\includegraphics[%
  width=0.35\columnwidth]{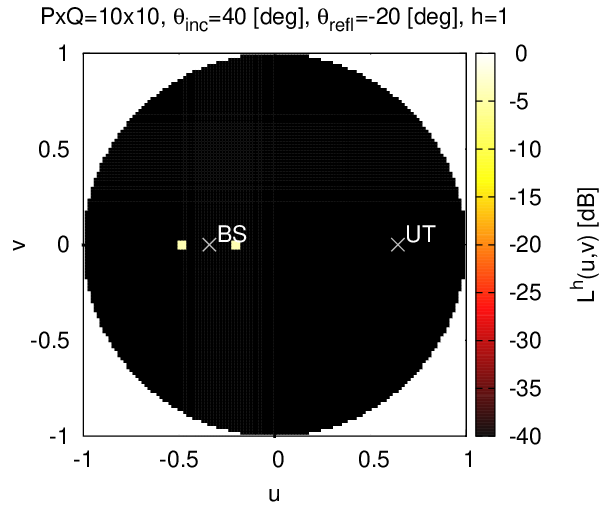}\tabularnewline
(\emph{a})&
(\emph{b})\tabularnewline
\includegraphics[%
  width=0.35\columnwidth]{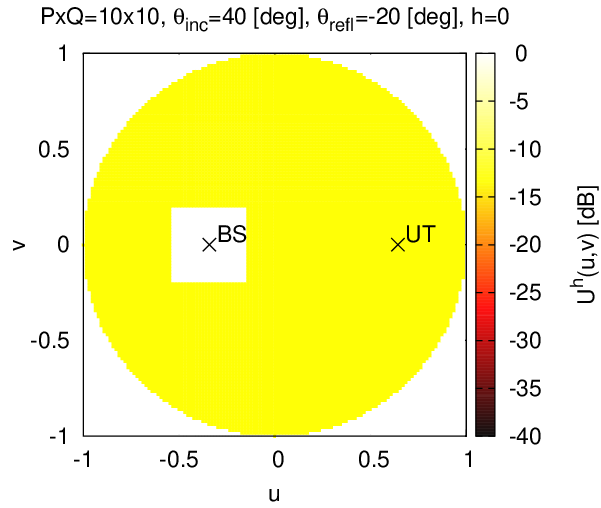}&
\includegraphics[%
  width=0.35\columnwidth]{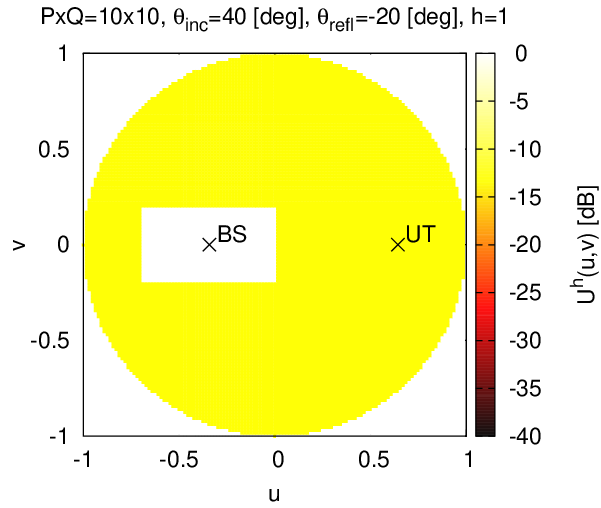}\tabularnewline
(\emph{c})&
(\emph{d})\tabularnewline
\includegraphics[%
  width=0.35\columnwidth]{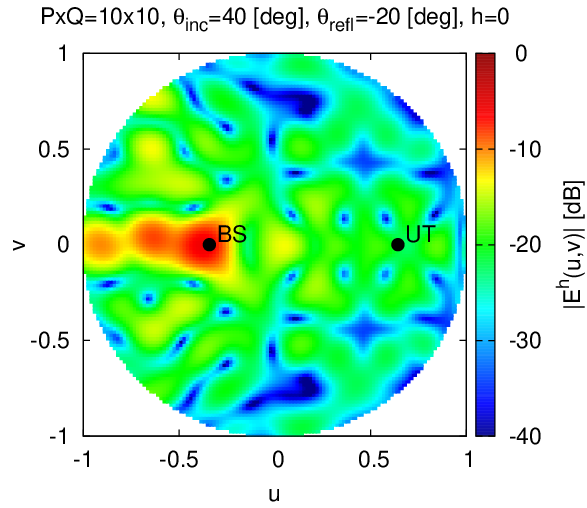}&
\includegraphics[%
  width=0.35\columnwidth]{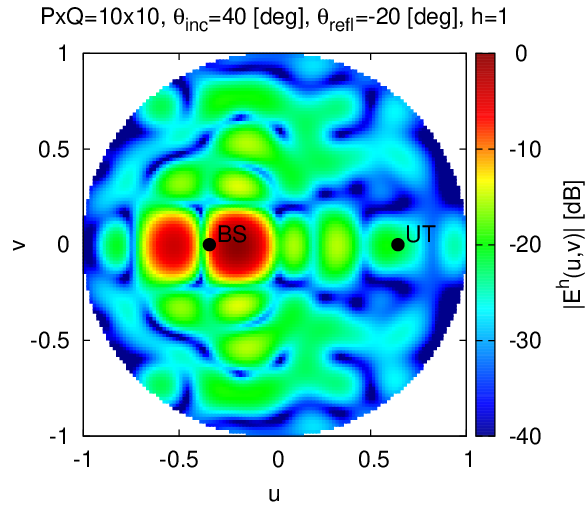}\tabularnewline
(\emph{e})&
(\emph{f})\tabularnewline
\includegraphics[%
  width=0.35\columnwidth]{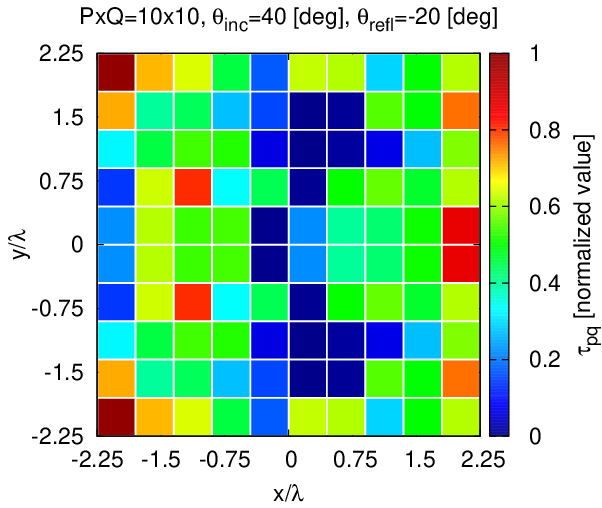}&
\includegraphics[%
  width=0.35\columnwidth]{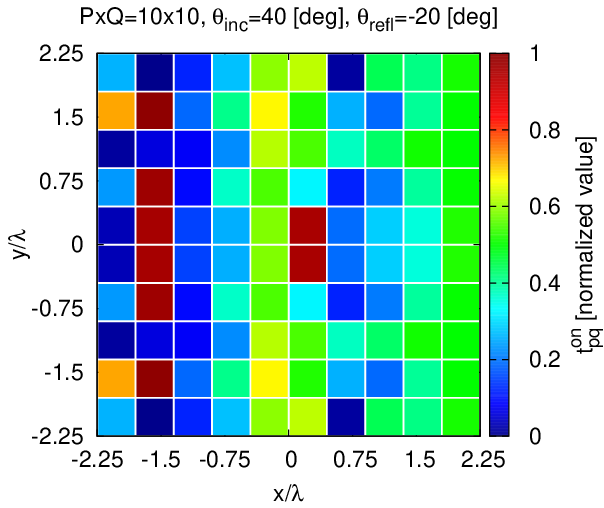}\tabularnewline
(\emph{g})&
(\emph{h})\tabularnewline
\end{tabular}\end{center}

\begin{center}\textbf{Fig. 2 - L. Poli} \textbf{\emph{et al.,}} {}``Time-Modulated
EM Skins for Integrated ...''\end{center}

\newpage
\begin{center}~\vfill\end{center}

\begin{center}\begin{tabular}{ccc}
\begin{sideways}
~~~~~~~~~$\theta_{refl}=-10$ {[}deg{]}%
\end{sideways}&
\includegraphics[%
  width=0.45\columnwidth]{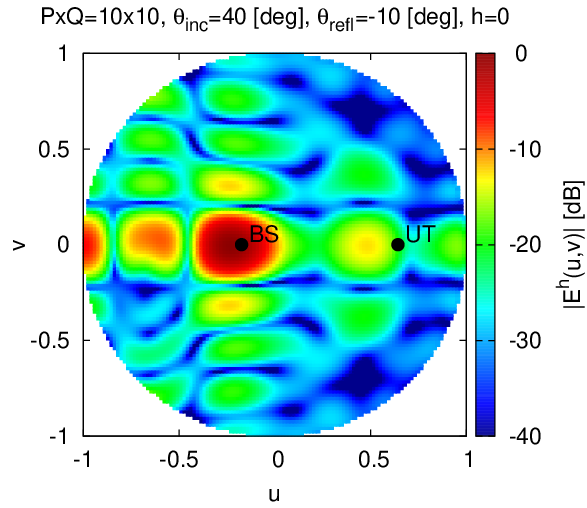}&
\includegraphics[%
  width=0.45\columnwidth]{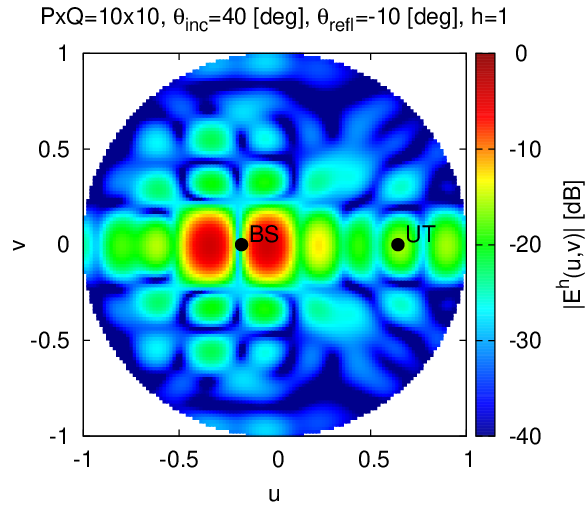}\tabularnewline
&
(\emph{a})&
(\emph{b})\tabularnewline
\begin{sideways}
~~~~~~~~~$\theta_{refl}=0$ {[}deg{]}%
\end{sideways}&
\includegraphics[%
  width=0.45\columnwidth]{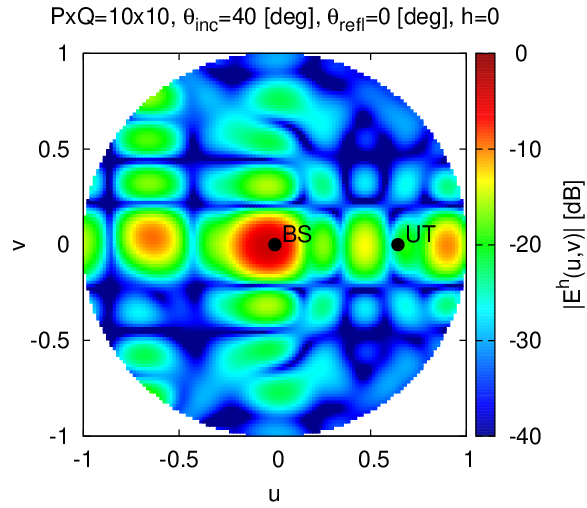}&
\includegraphics[%
  width=0.45\columnwidth]{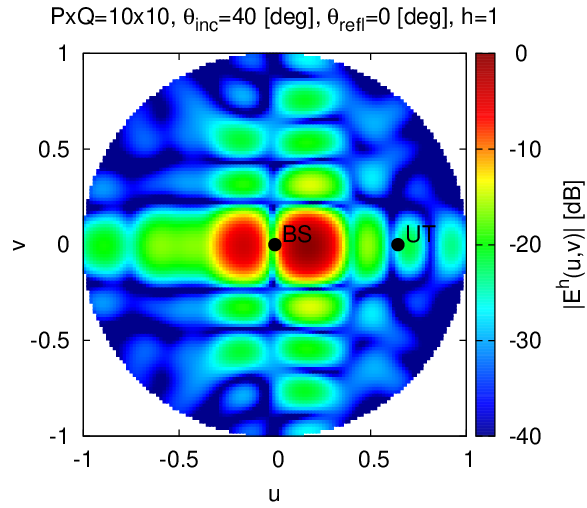}\tabularnewline
&
(\emph{c})&
(\emph{d})\tabularnewline
\end{tabular}\end{center}

\begin{center}~\vfill\end{center}

\begin{center}\textbf{Fig. 3 - L. Poli} \textbf{\emph{et al.,}} {}``Time-Modulated
EM Skins for Integrated ...''\end{center}

\newpage
\begin{center}~\vfill\end{center}

\begin{center}\includegraphics[%
  width=0.90\columnwidth]{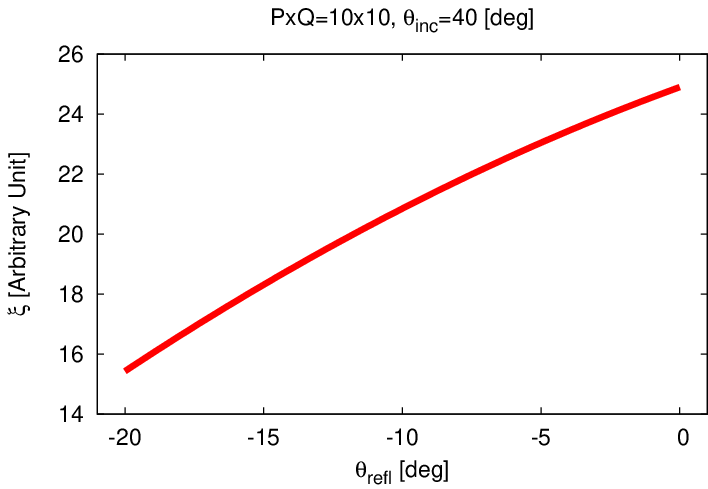}\end{center}

\begin{center}~\vfill\end{center}

\begin{center}\textbf{Fig. 4 - L. Poli} \textbf{\emph{et al.,}} {}``Time-Modulated
EM Skins for Integrated ...''\end{center}

\newpage
\begin{center}~\vfill\end{center}

\begin{center}\begin{tabular}{cc}
\multicolumn{2}{c}{\includegraphics[%
  width=0.70\columnwidth]{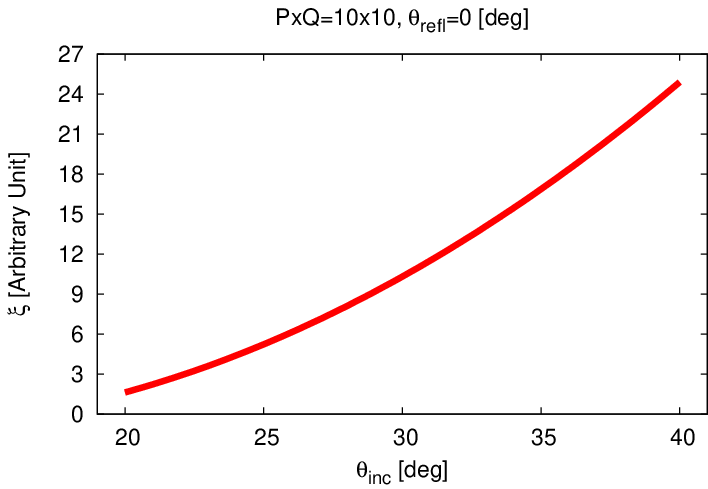}}\tabularnewline
\multicolumn{2}{c}{(\emph{a})}\tabularnewline
\includegraphics[%
  width=0.40\columnwidth]{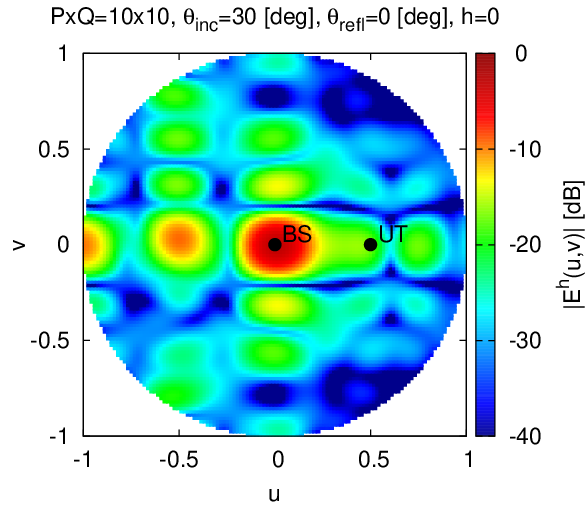}&
\includegraphics[%
  width=0.40\columnwidth]{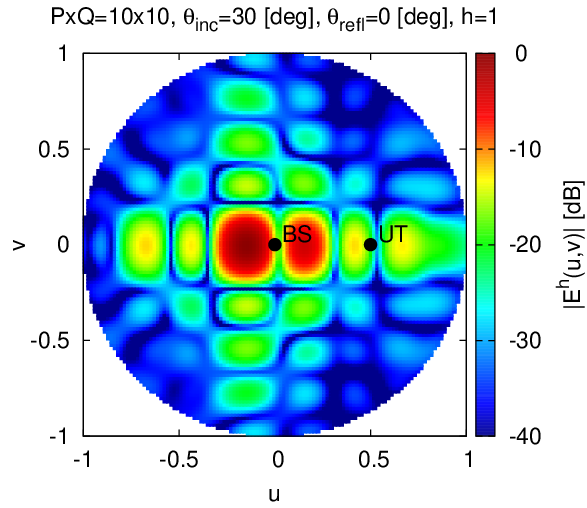}\tabularnewline
(\emph{b})&
(\emph{c})\tabularnewline
\includegraphics[%
  width=0.40\columnwidth]{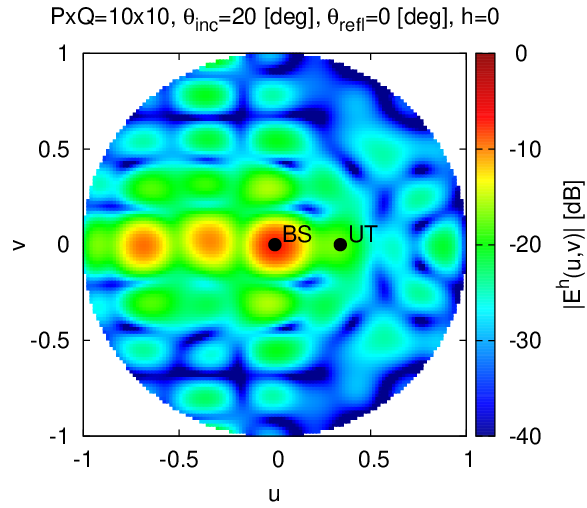}&
\includegraphics[%
  width=0.40\columnwidth]{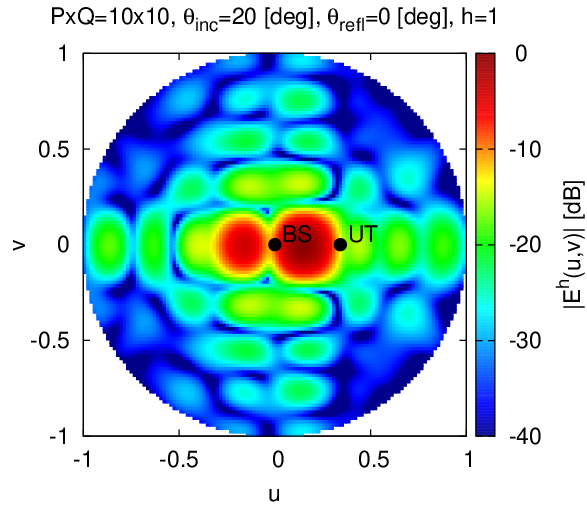}\tabularnewline
(\emph{d})&
(\emph{e})\tabularnewline
\end{tabular}\end{center}

\begin{center}~\vfill\end{center}

\begin{center}\textbf{Fig. 5 - L. Poli} \textbf{\emph{et al.,}} {}``Time-Modulated
EM Skins for Integrated ...''\end{center}

\newpage
\begin{center}~\vfill\end{center}

\begin{center}\begin{tabular}{cc}
\multicolumn{2}{c}{\includegraphics[%
  width=0.70\columnwidth]{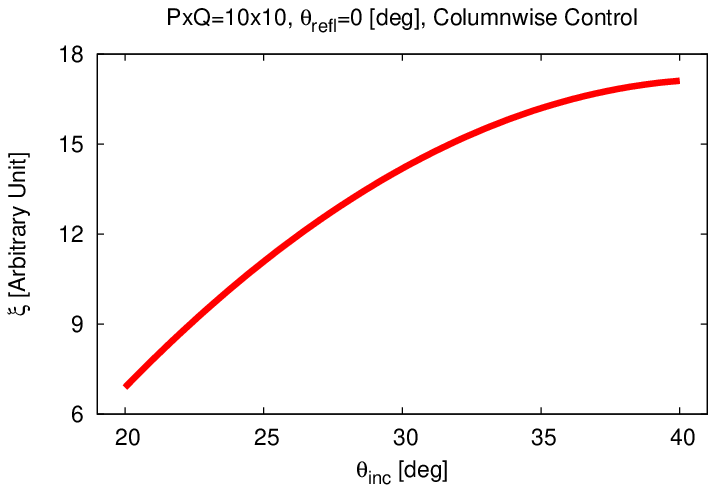}}\tabularnewline
\multicolumn{2}{c}{(\emph{a})}\tabularnewline
\includegraphics[%
  width=0.40\columnwidth]{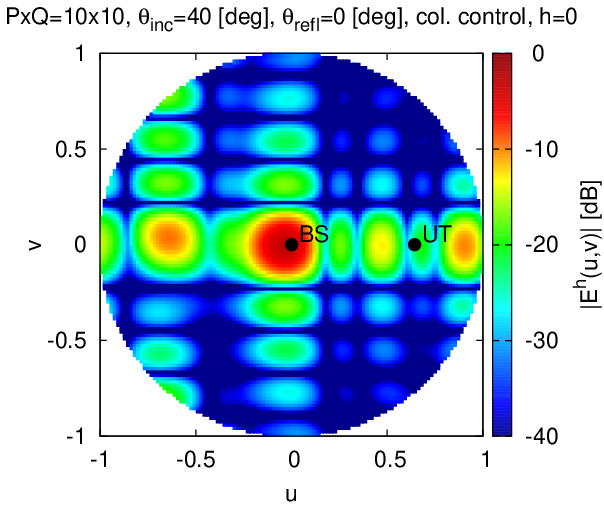}&
\includegraphics[%
  width=0.40\columnwidth]{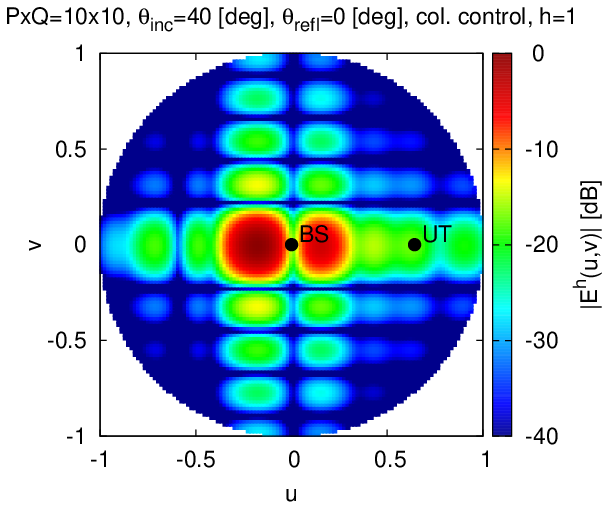}\tabularnewline
(\emph{b})&
(\emph{c})\tabularnewline
\includegraphics[%
  width=0.40\columnwidth]{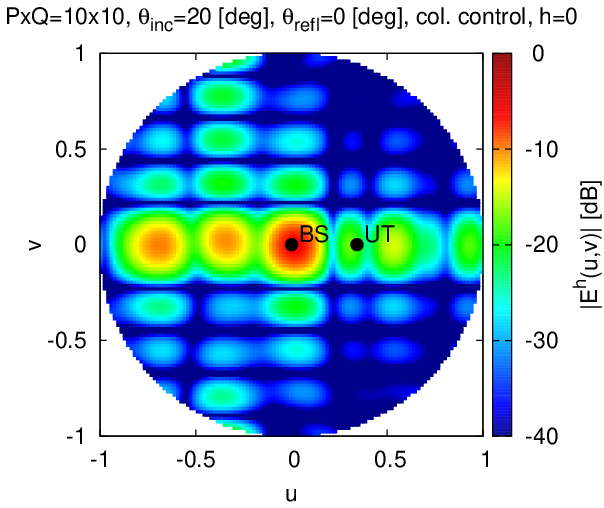}&
\includegraphics[%
  width=0.40\columnwidth]{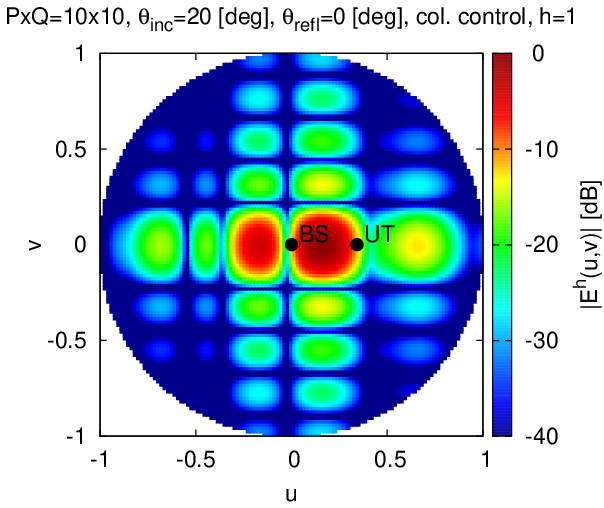}\tabularnewline
(\emph{d})&
(\emph{e})\tabularnewline
\end{tabular}\end{center}

\begin{center}~\vfill\end{center}

\begin{center}\textbf{Fig. 6 - L. Poli} \textbf{\emph{et al.,}} {}``Time-Modulated
EM Skins for Integrated ...''\end{center}

\newpage
\begin{center}~\vfill\end{center}

\begin{center}\begin{tabular}{cc}
\includegraphics[%
  width=0.50\columnwidth]{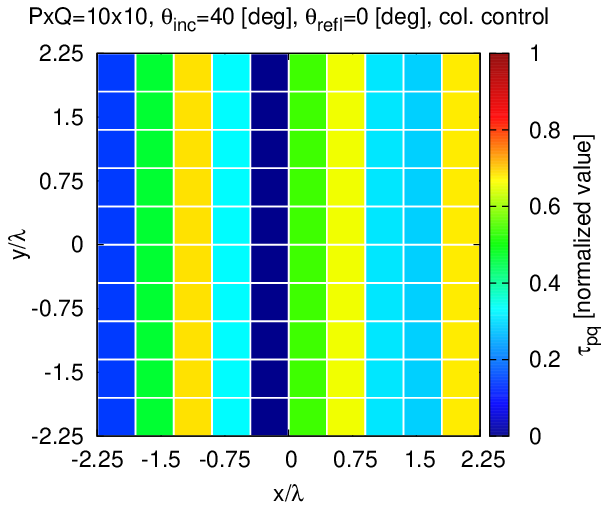}&
\includegraphics[%
  width=0.50\columnwidth]{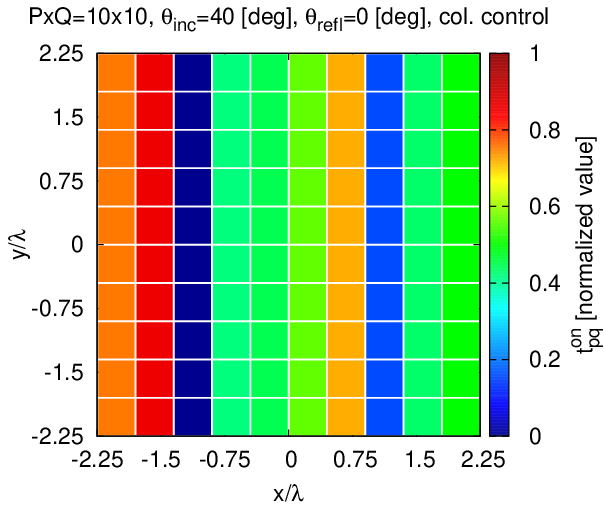}\tabularnewline
(\emph{a})&
(\emph{b})\tabularnewline
\includegraphics[%
  width=0.50\columnwidth]{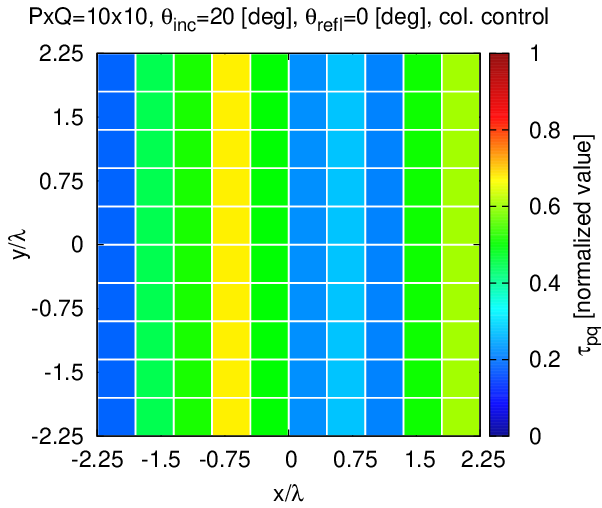}&
\includegraphics[%
  width=0.50\columnwidth]{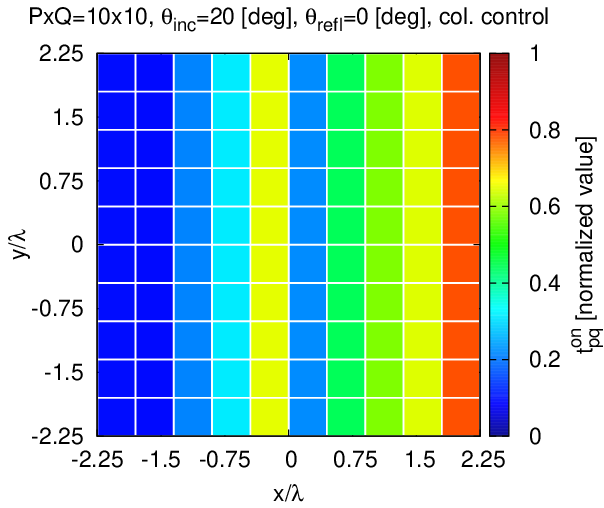}\tabularnewline
(\emph{c})&
(\emph{d})\tabularnewline
\end{tabular}\end{center}

\begin{center}~\vfill\end{center}

\begin{center}\textbf{Fig. 7 - L. Poli} \textbf{\emph{et al.,}} {}``Time-Modulated
EM Skins for Integrated ...''\end{center}

\newpage
\begin{center}~\vfill\end{center}

\begin{center}\begin{tabular}{ccccc}
\multicolumn{5}{c}{\includegraphics[%
  width=0.60\columnwidth]{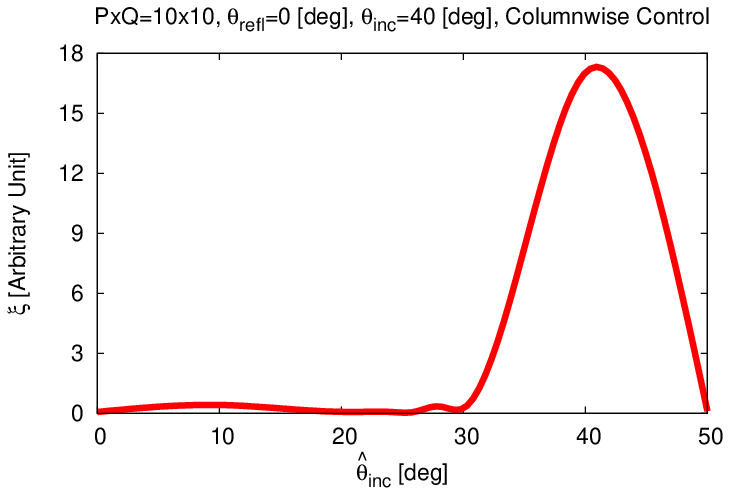}}\tabularnewline
\multicolumn{5}{c}{(\emph{a})}\tabularnewline
\begin{sideways}
~~~~~$\widehat{\theta}_{inc}=0$ {[}deg{]}%
\end{sideways}&
\includegraphics[%
  width=0.21\columnwidth]{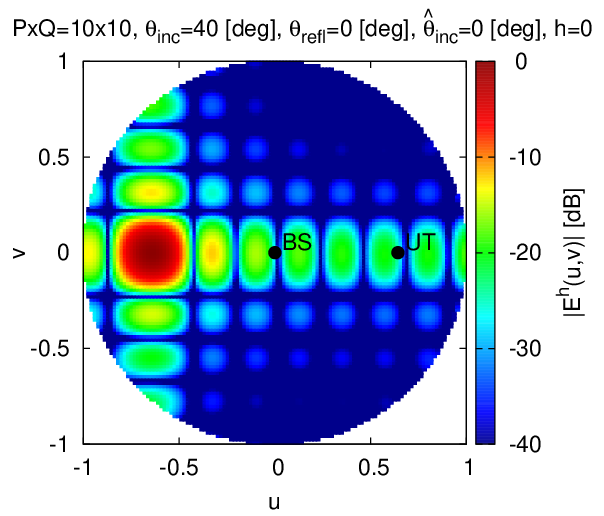}&
\includegraphics[%
  width=0.21\columnwidth]{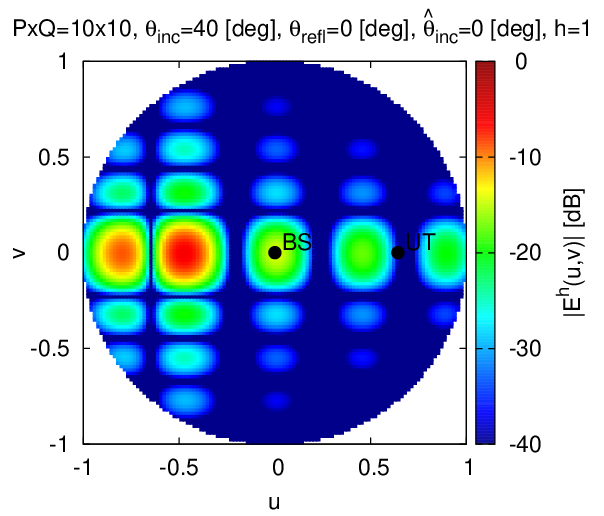}&
\includegraphics[%
  width=0.21\columnwidth]{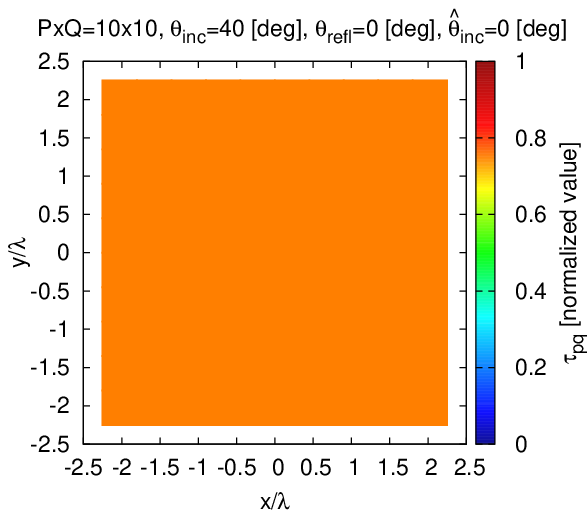}&
\includegraphics[%
  width=0.21\columnwidth]{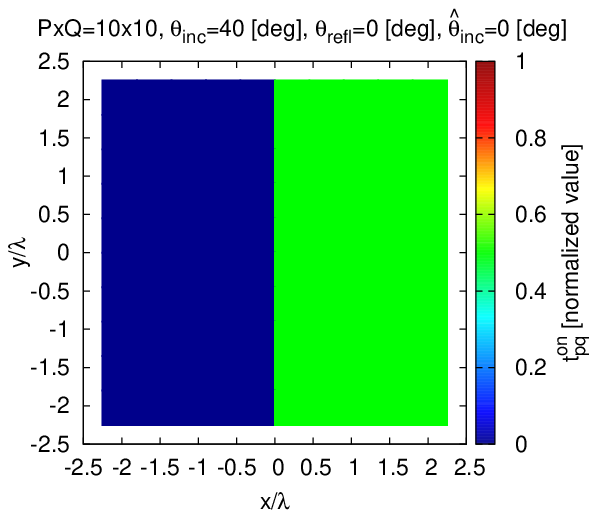}\tabularnewline
&
(\emph{b})&
(\emph{c})&
(\emph{d})&
(\emph{e})\tabularnewline
\begin{sideways}
~~~~~$\widehat{\theta}_{inc}=20$ {[}deg{]}%
\end{sideways}&
\includegraphics[%
  width=0.21\columnwidth]{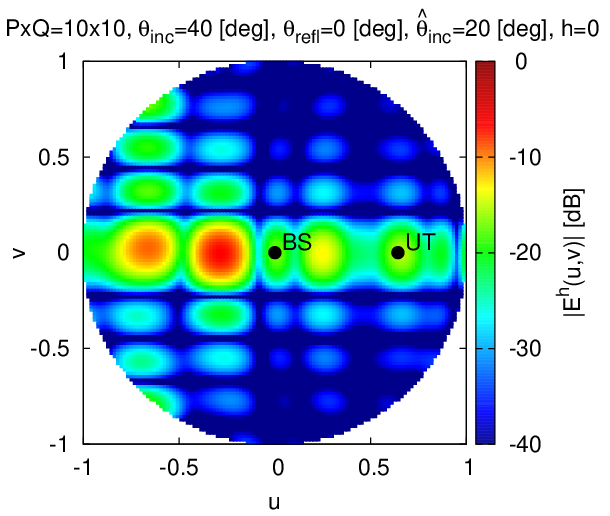}&
\includegraphics[%
  width=0.21\columnwidth]{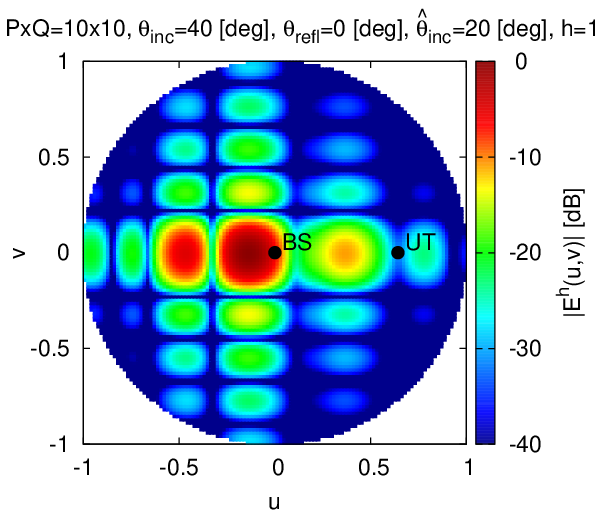}&
\includegraphics[%
  width=0.21\columnwidth]{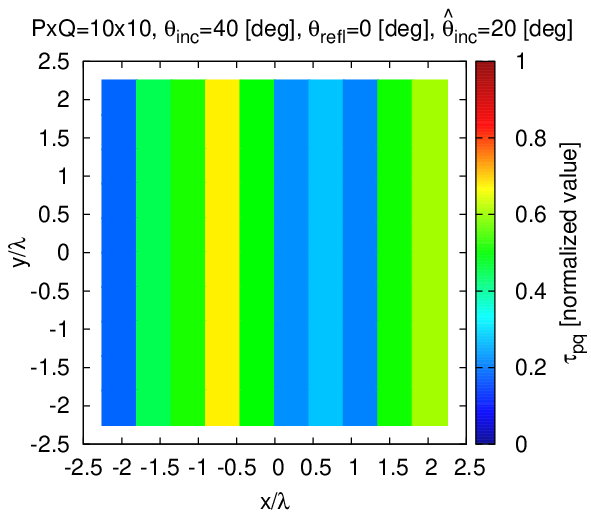}&
\includegraphics[%
  width=0.21\columnwidth]{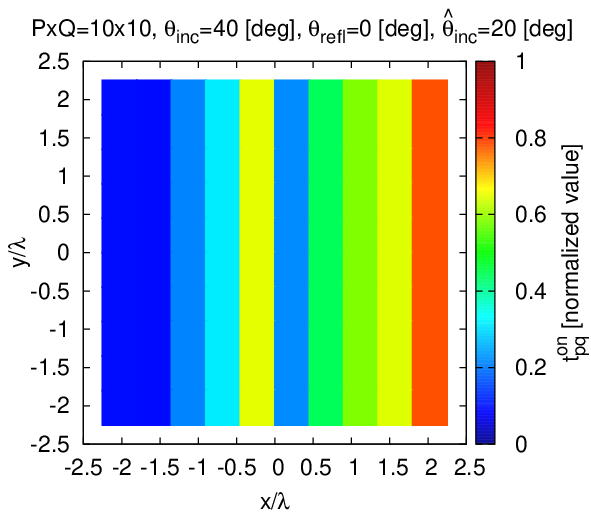}\tabularnewline
&
(\emph{f})&
(\emph{g})&
(\emph{h})&
(\emph{i})\tabularnewline
\begin{sideways}
~~~~~$\widehat{\theta}_{inc}=40$ {[}deg{]}%
\end{sideways}&
\includegraphics[%
  width=0.21\columnwidth]{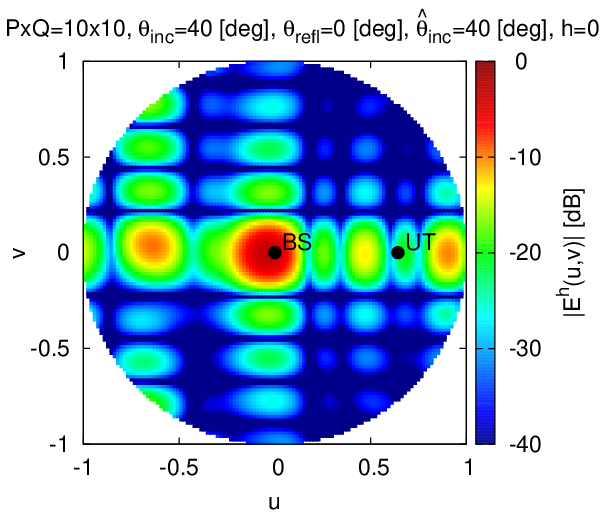}&
\includegraphics[%
  width=0.21\columnwidth]{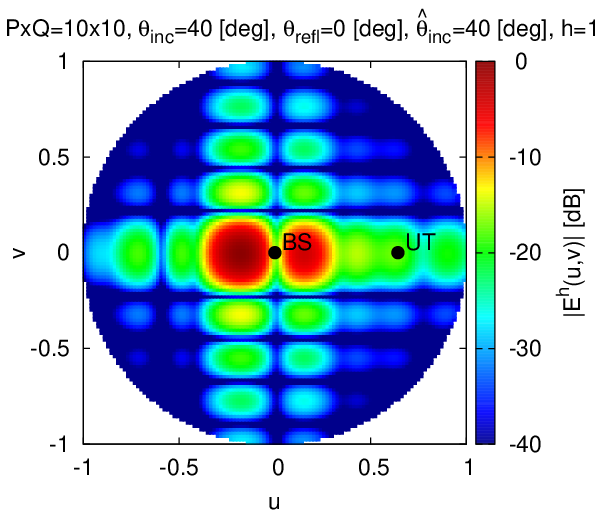}&
\includegraphics[%
  width=0.21\columnwidth]{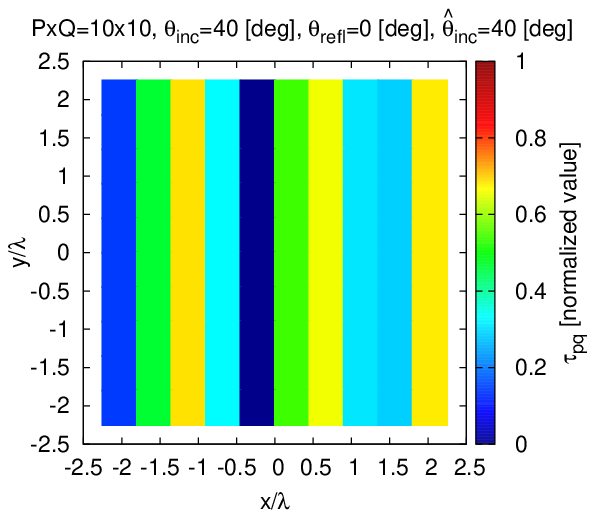}&
\includegraphics[%
  width=0.21\columnwidth]{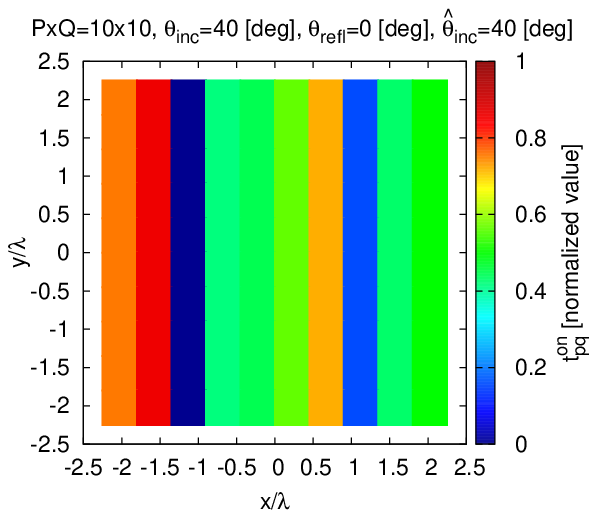}\tabularnewline
&
(\emph{j})&
(\emph{k})&
(\emph{l})&
(\emph{m})\tabularnewline
\end{tabular}\end{center}

\begin{center}~\vfill\end{center}

\begin{center}\textbf{Fig. 8 - L. Poli} \textbf{\emph{et al.,}} {}``Time-Modulated
EM Skins for Integrated ...''\end{center}

\newpage
\begin{center}~\vfill\end{center}

\begin{center}\begin{tabular}{cc}
\includegraphics[%
  width=0.50\columnwidth]{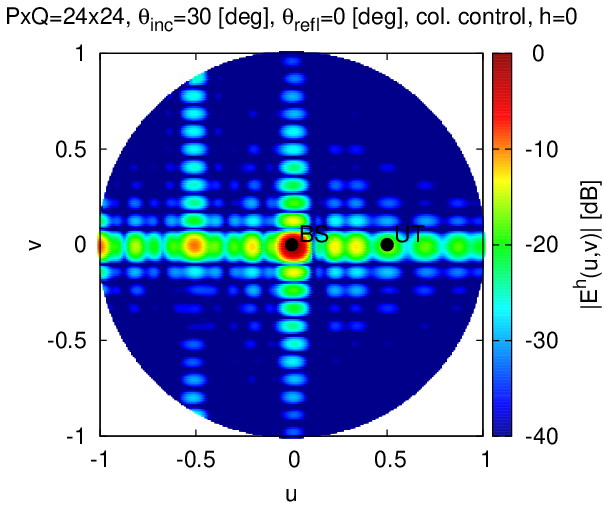}&
\includegraphics[%
  width=0.50\columnwidth]{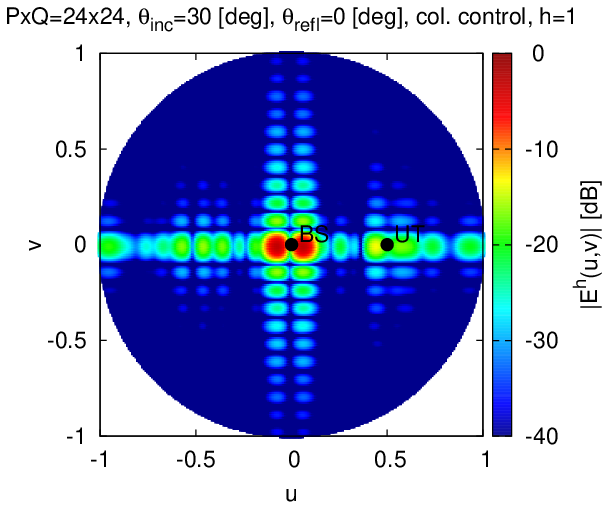}\tabularnewline
(\emph{a})&
(\emph{b})\tabularnewline
\includegraphics[%
  width=0.50\columnwidth]{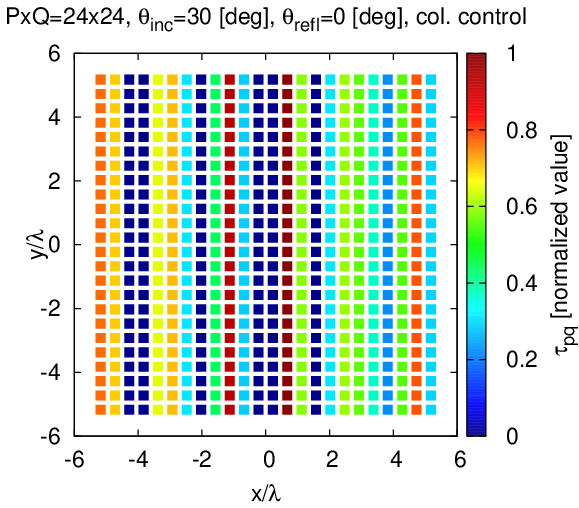}&
\includegraphics[%
  width=0.50\columnwidth]{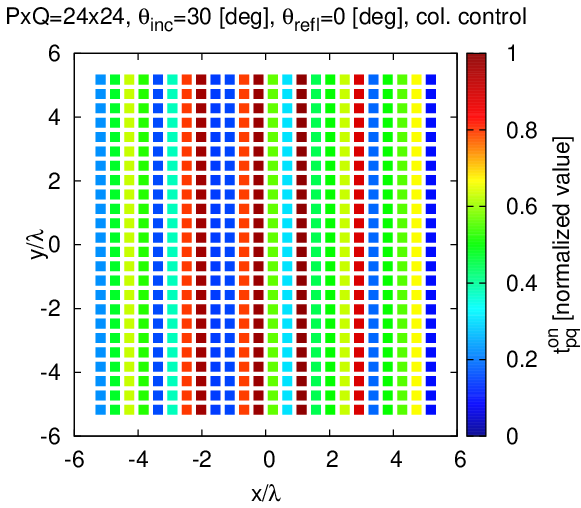}\tabularnewline
(\emph{c})&
(\emph{d})\tabularnewline
\end{tabular}\end{center}

\begin{center}~\vfill\end{center}

\begin{center}\textbf{Fig. 9 - L. Poli} \textbf{\emph{et al.,}} {}``Time-Modulated
EM Skins for Integrated ...''\end{center}

\newpage
\begin{center}~\vfill\end{center}

\begin{center}\begin{tabular}{c}
\includegraphics[%
  width=0.40\columnwidth]{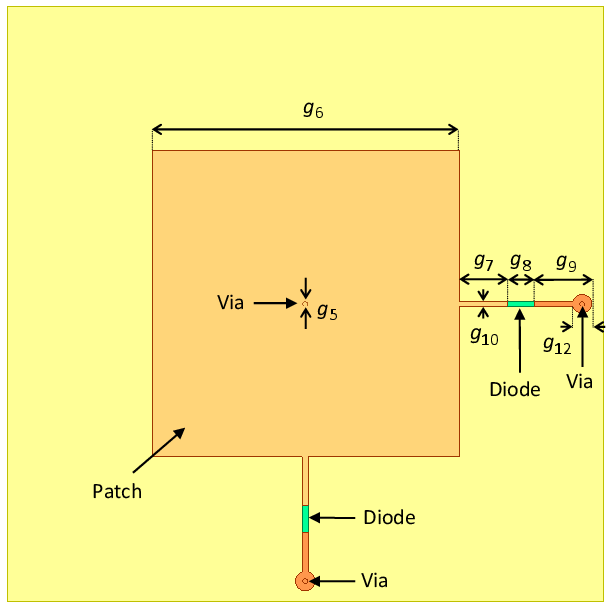}\tabularnewline
(\emph{a})\tabularnewline
\includegraphics[%
  width=0.40\columnwidth]{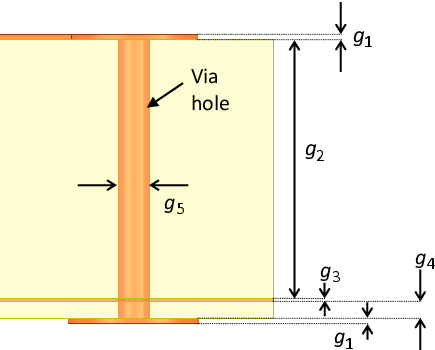}\tabularnewline
(\emph{b})\tabularnewline
\includegraphics[%
  width=0.40\columnwidth]{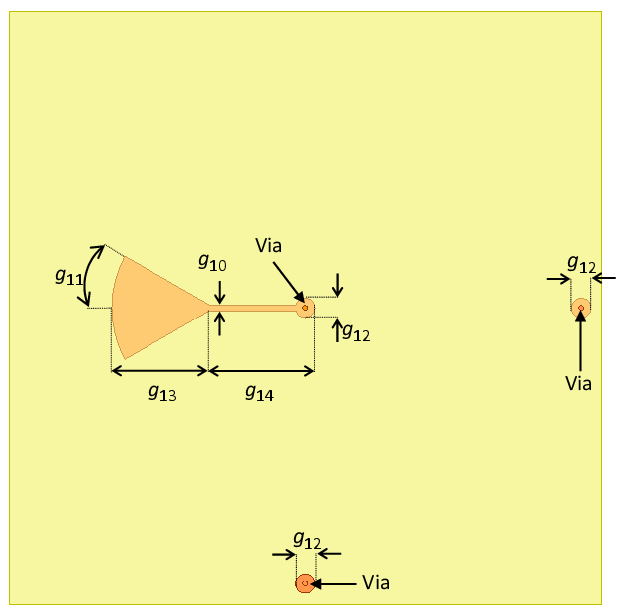}\tabularnewline
(\emph{c})\tabularnewline
\end{tabular}\end{center}

\begin{center}~\vfill\end{center}

\begin{center}\textbf{Fig. 10 - L. Poli} \textbf{\emph{et al.,}} {}``Time-Modulated
EM Skins for Integrated ...''\end{center}

\newpage
\begin{center}~\end{center}

\begin{center}\begin{tabular}{cc}
\multicolumn{2}{c}{\includegraphics[%
  width=0.60\columnwidth]{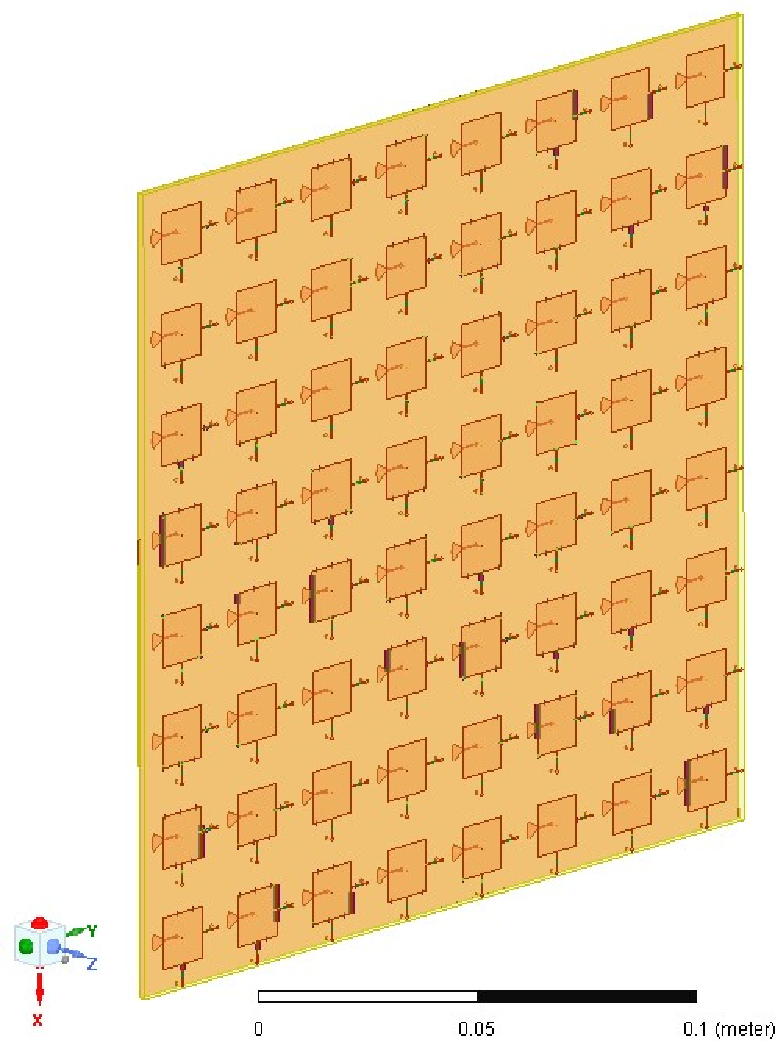}}\tabularnewline
\multicolumn{2}{c}{(\emph{a})}\tabularnewline
\includegraphics[%
  width=0.50\columnwidth]{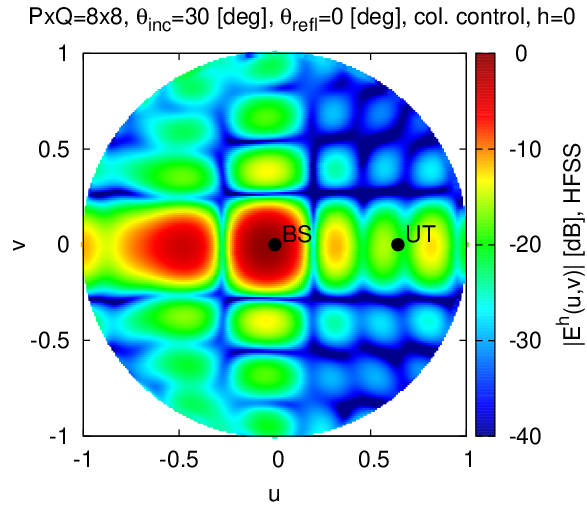}&
\includegraphics[%
  width=0.50\columnwidth]{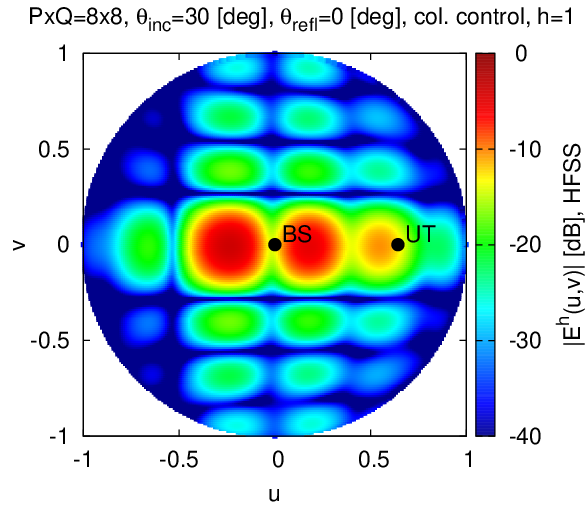}\tabularnewline
(\emph{b})&
(\emph{c})\tabularnewline
\end{tabular}\end{center}

\begin{center}~\end{center}

\begin{center}\textbf{Fig. 11 - L. Poli} \textbf{\emph{et al.,}} {}``Time-Modulated
EM Skins for Integrated ...''\end{center}

\newpage
\begin{center}\begin{tabular}{cc}
\multicolumn{2}{c}{\includegraphics[%
  width=0.58\columnwidth]{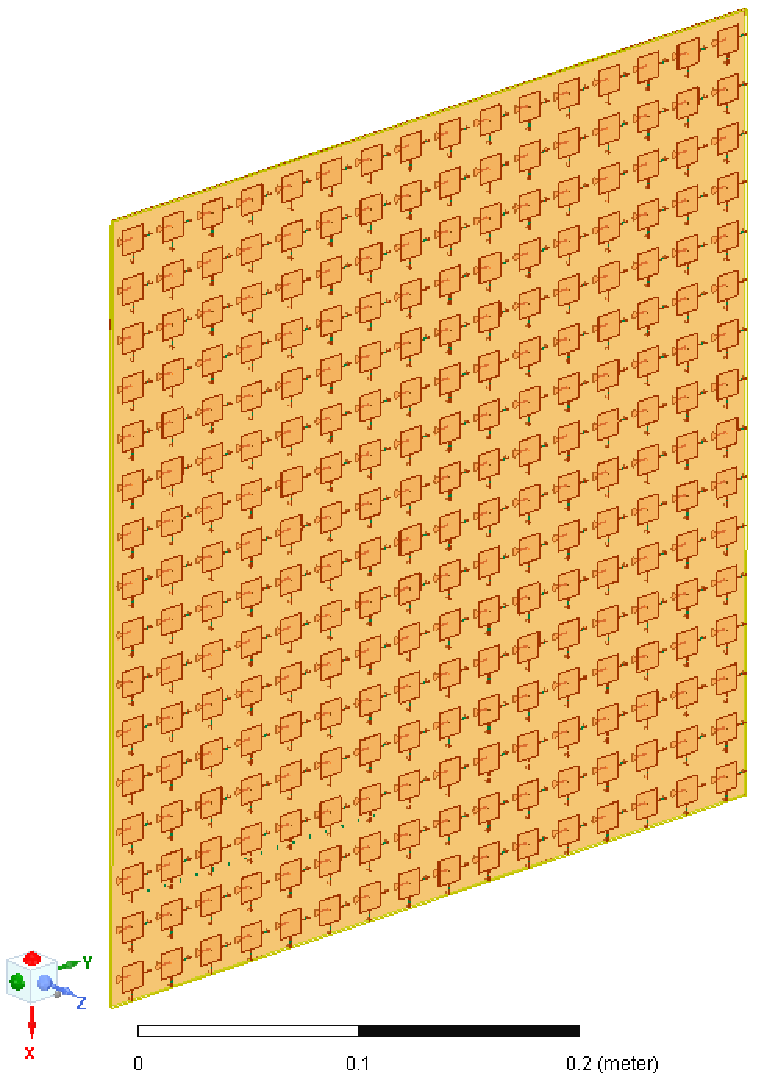}}\tabularnewline
\multicolumn{2}{c}{(\emph{a})}\tabularnewline
\includegraphics[%
  width=0.50\columnwidth]{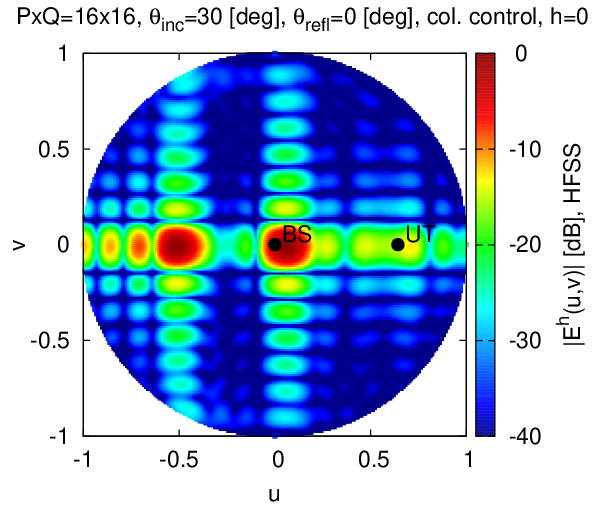}&
\includegraphics[%
  width=0.50\columnwidth]{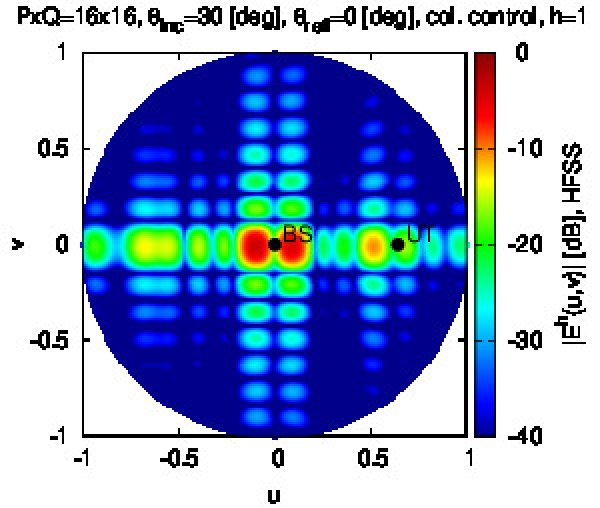}\tabularnewline
(\emph{b})&
(\emph{c})\tabularnewline
\end{tabular}\end{center}

\begin{center}~\end{center}

\begin{center}\textbf{Fig. 12 - L. Poli} \textbf{\emph{et al.,}} {}``Time-Modulated
EM Skins for Integrated ...''\end{center}

\newpage
\begin{center}~\vfill\end{center}

\begin{center}\begin{tabular}{cc}
\includegraphics[%
  width=0.45\columnwidth]{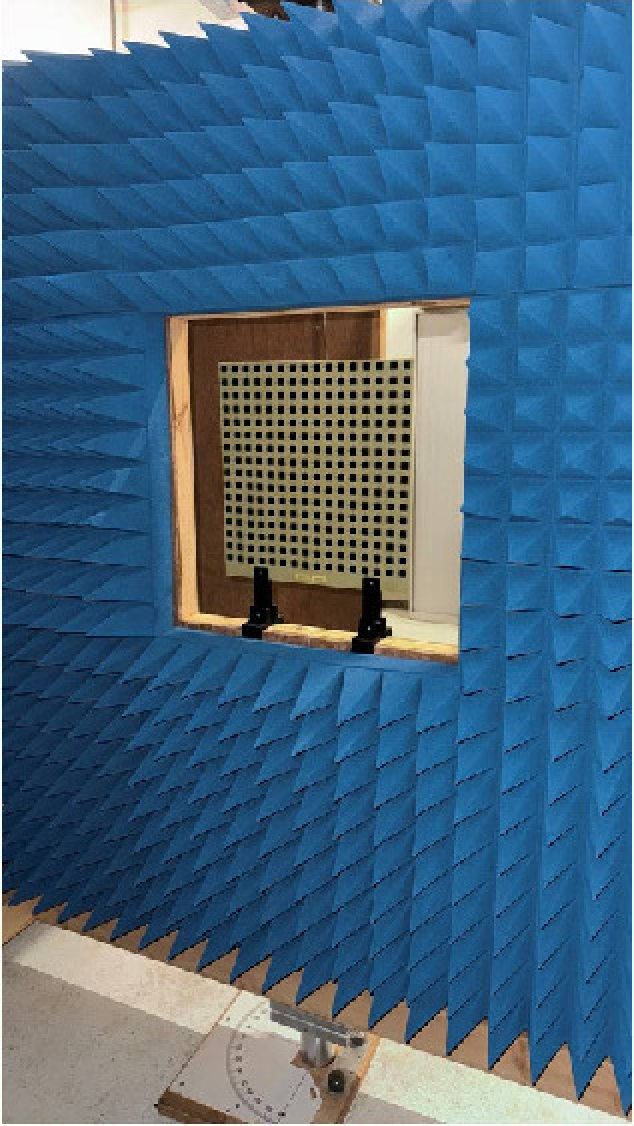}&
\includegraphics[%
  width=0.45\columnwidth]{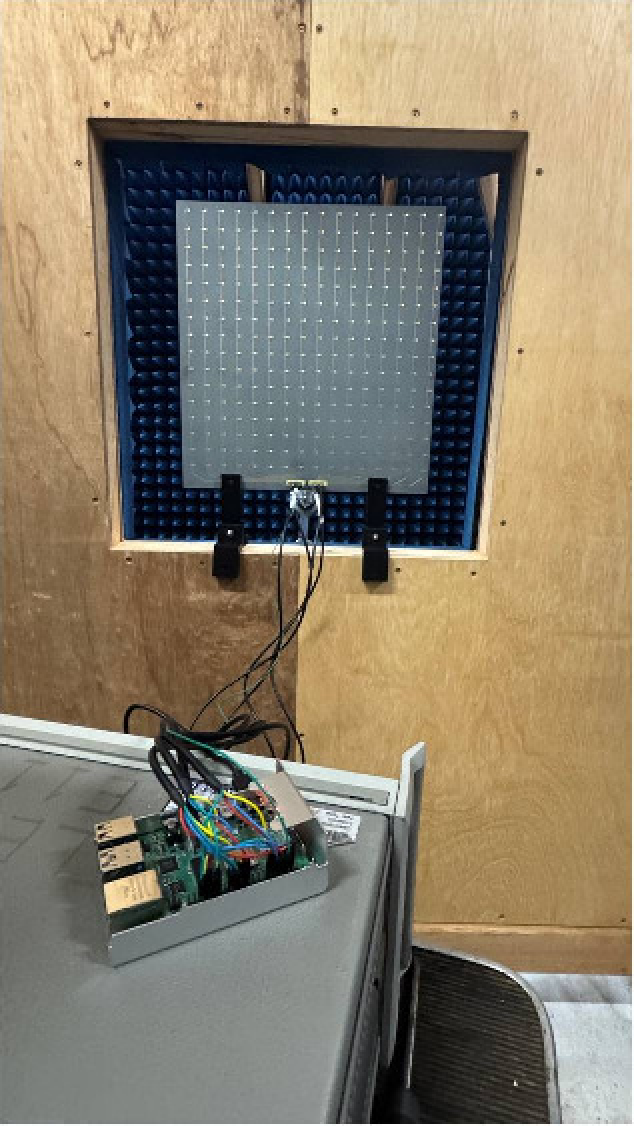}\tabularnewline
(\emph{a})&
(\emph{b})\tabularnewline
\end{tabular}\end{center}

\begin{center}\vfill~\end{center}

\begin{center}\textbf{Fig. 13 - L. Poli} \textbf{\emph{et al.,}} {}``Time-Modulated
EM Skins for Integrated ...''\end{center}

\newpage
\begin{center}~\vfill\end{center}

\begin{center}\begin{tabular}{cc}
\multicolumn{2}{c}{\includegraphics[%
  width=0.95\columnwidth]{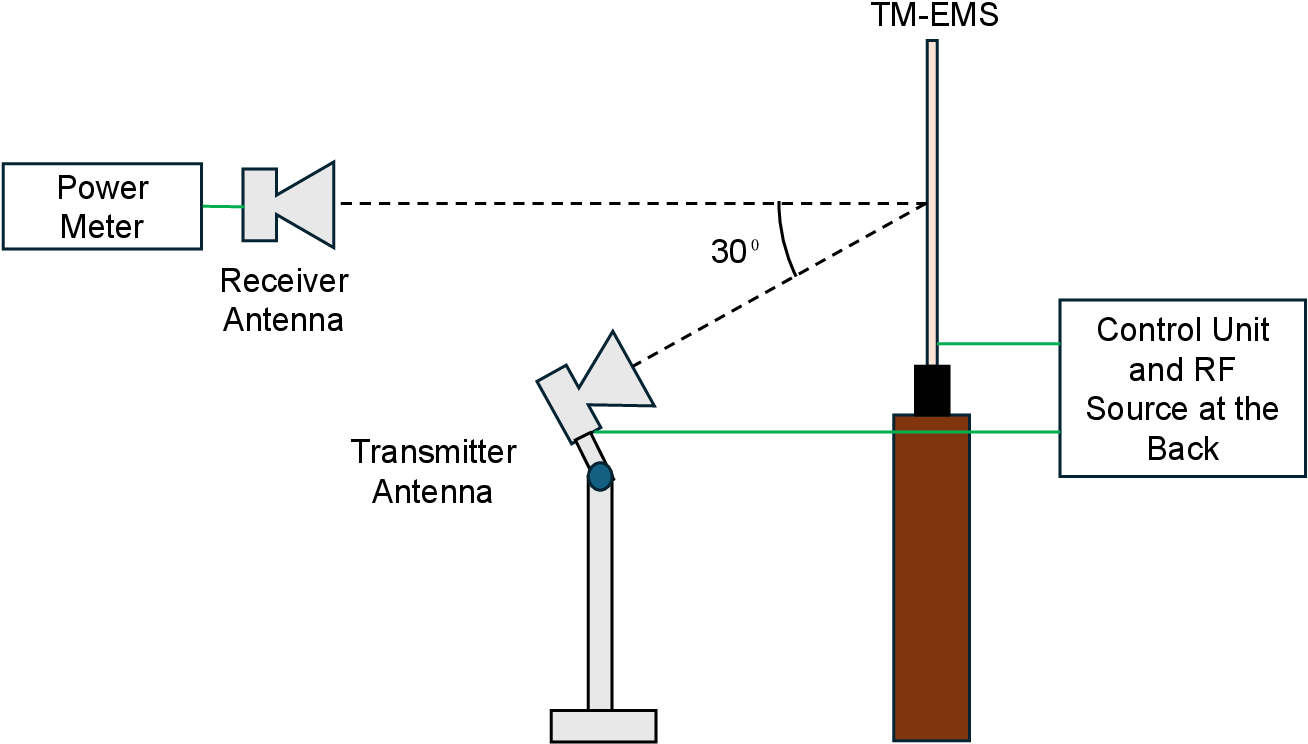}}\tabularnewline
\multicolumn{2}{c}{(\emph{a})}\tabularnewline
\multicolumn{2}{c}{\includegraphics[%
  width=0.95\columnwidth]{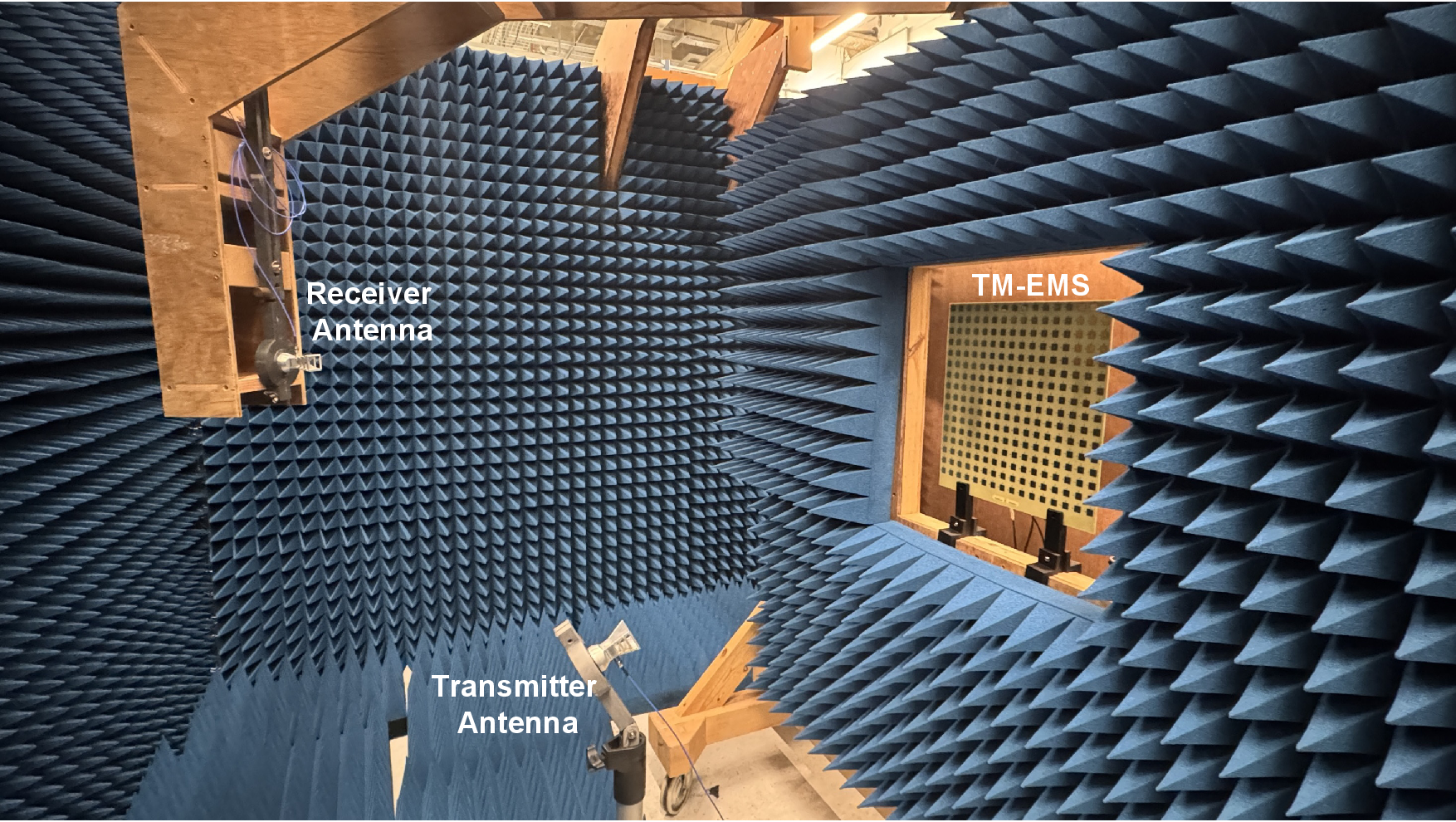}}\tabularnewline
\multicolumn{2}{c}{(\emph{b})}\tabularnewline
\end{tabular}\end{center}

\begin{center}\vfill~\end{center}

\begin{center}\textbf{Fig. 14 - L. Poli} \textbf{\emph{et al.,}} {}``Time-Modulated
EM Skins for Integrated ...''\end{center}

\newpage
\begin{center}~\vfill\end{center}

\begin{center}\begin{tabular}{c}
\includegraphics[%
  width=0.90\columnwidth]{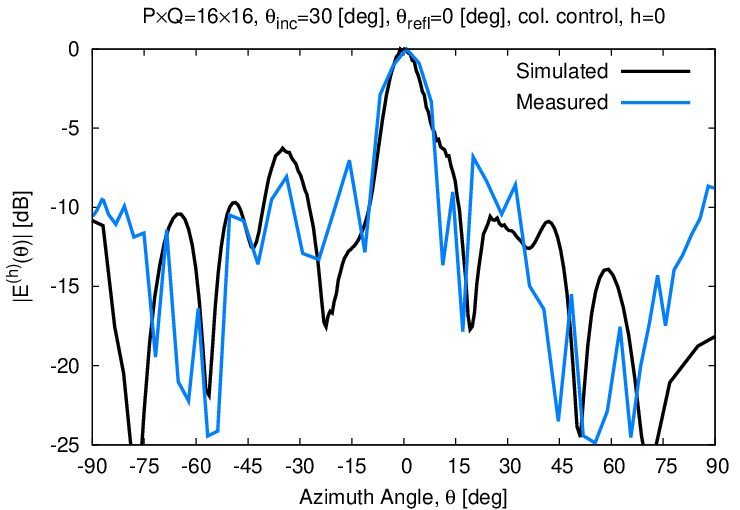}\tabularnewline
(\emph{a})\tabularnewline
\includegraphics[%
  width=0.90\columnwidth]{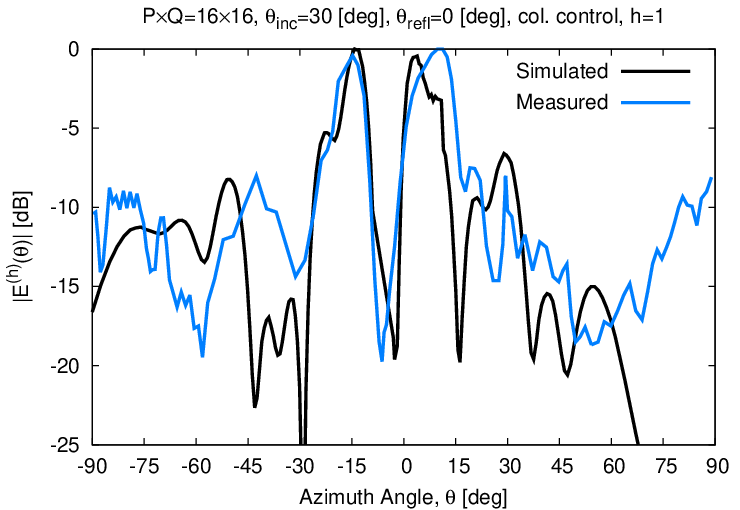}\tabularnewline
(\emph{b})\tabularnewline
\end{tabular}\end{center}

\begin{center}~\vfill\end{center}

\begin{center}\textbf{Fig. 15 - L. Poli} \textbf{\emph{et al.,}} {}``Time-Modulated
EM Skins for Integrated ...''\end{center}

\newpage
\begin{center}~\vfill\end{center}

\begin{center}\begin{tabular}{|c|c|c|c|}
\hline 
\textbf{Parameter}&
\textbf{Value}&
\textbf{Parameter}&
\textbf{Value}\tabularnewline
\hline
\hline 
$g_{1}$&
$35$ {[}$\mu$m{]}&
$g_{8}$&
$1.1$ {[}mm{]}\tabularnewline
\hline 
$g_{2}$&
$1.6$ {[}mm{]}&
$g_{9}$&
$2.4$ {[}mm{]}\tabularnewline
\hline 
$g_{3}$&
$15.2$ {[}$\mu$m{]}&
$g_{10}$&
$0.3$ {[}mm{]}\tabularnewline
\hline 
$g_{4}$&
$0.1$ {[}mm{]}&
$g_{11}$&
$30$ {[}deg{]}\tabularnewline
\hline 
$g_{5}$&
$0.2$ {[}mm{]}&
$g_{12}$&
$0.8$ {[}mm{]}\tabularnewline
\hline 
$g_{6}$&
$12.6$ {[}mm{]}&
$g_{13}$&
$4.0$ {[}mm{]}\tabularnewline
\hline 
$g_{7}$&
$2.0$ {[}mm{]}&
$g_{14}$&
$4.4$ {[}mm{]}\tabularnewline
\hline
\end{tabular}\end{center}

\begin{center}~\vfill\end{center}

\begin{center}\textbf{Table I - L. Poli} \textbf{\emph{et al.,}} {}``Time-Modulated
EM Skins for Integrated ...''\end{center}
\end{document}